\documentclass[preprint,showpacs,amsmath,floatfix,prb,aps]{revtex4}

\usepackage{color}
\usepackage{amsmath}
\usepackage{pifont}  
\usepackage{graphicx}
\usepackage{dcolumn} 
\usepackage{bm}      
\usepackage{amsfonts}
\usepackage{amssymb} 

\newcommand{\ba}{{\bf a}}
\newcommand{\bb}{{\bf b}}

\newcommand{\bg}{{\bf g}}

\newcommand{\bk}{{\bf k}}
\newcommand{\bl}{{\bf l}}
\newcommand{\bnn}{{\bf n}}
\newcommand{\bem}{{\bf m}}
\newcommand{\bpp}{{\bf p}}
\newcommand{\br}{{\bf r}}

\newcommand{\bt}{{\bf t}}
\newcommand{\bz}{{\bf z}}

\newcommand{\bG}{{\bf G}}
\newcommand{\bH}{{\bf H}}
\newcommand{\bK}{{\bf K}}

\newcommand{\bT}{{\bf T}}
\newcommand{\bR}{{\bf R}}
\newcommand{\bU}{{\bf U}}
\newcommand{\bV}{{\bf V}}

\newcommand{\GCD}{\text{gcd}}

\newcommand{\tH}{\text{H}}

\newcommand{\kh}[2]{\ensuremath{ \left|\phi_{#1\bk}^{(#2)}\right\rangle }}

\newcommand{\ki}[2]{\ensuremath{ \left|\phi_{#1}^{(#2)}\right\rangle }}
\newcommand{\bi}[2]{\ensuremath{ \left\langle\phi_{#1}^{(#2)}\right| }}

\newcommand{\kn}[1]{\ensuremath{ \left|\phi_{i_{#1}\bk_{#1}}^{(#1)}\right\rangle }}
\newcommand{\bn}[1]{\ensuremath{ \left\langle\phi_{i_{#1}\bk_{#1}}^{(#1)}\right| }}
\newcommand{\prdn}[2]{\ensuremath{ 
 \left\langle\phi_{i_#1\bk_#1}^{(#1)}|\phi_{i_{#2}\bk_{#2}}^{(#2)}\right\rangle }}

\newcommand{\ep}[2]{\ensuremath{ \epsilon_{#1\bk}^{(#2)} }}

\newcommand{\gk}[1]{\left|#1\right\rangle}
\newcommand{\gb}[1]{\left\langle#1\right|}

\begin{document}

\author{S. Shallcross$^{1}$}
\email{phsss@tfkp.physik.uni-erlangen.de}
\author{S. Sharma$^{2}$}
\author{E. Kandelaki$^{1,3}$}
\author{O.~A. Pankratov$^{1}$}
\affiliation{1 Lehrstuhl f\"ur Theoretische Festk\"orperphysik, Staudstr. 7-B2, 91058 Erlangen, Germany.}
\affiliation{2 Max-Planck Institute for Microstructure Physics, Weinberg 2, 06120 Halle, Germany.}
\affiliation{3 Ruhr-Universit\"at Bochum, Universit\"atsstrasse 150, 44801 Bochum, Germany.}

\title{Electronic structure of turbostratic graphene}

\begin{abstract}

We explore the rotational degree of freedom between graphene layers via
the simple prototype of the graphene twist bilayer, i.e., two 
layers rotated by some angle $\theta$. It is shown that, due to the weak
interaction between graphene layers, many features of this
system can be understood by interference conditions between
the quantum states of the two layers, mathematically expressed
as Diophantine problems. Based on this general analysis
we demonstrate that while the Dirac cones from each layer
are always effectively degenerate,
the Fermi velocity $v_F$ of the Dirac cones
decreases as $\theta\rightarrow 0^\circ$; the form we derive for $v_F(\theta)$
agrees with that found via a continuum approximation in
Ref.~[\onlinecite{san07}]. From tight binding calculations for structures
with $1.47^\circ \le \theta < 30^\circ$ we find agreement
with this formula for $\theta \gtrsim 5^\circ$.
In contrast, for $\theta \lesssim 5^\circ$ this formula breaks down
and the Dirac bands become strongly warped as
the limit $\theta \rightarrow 0$ is approached.
For an ideal system of twisted layers the limit as $\theta\rightarrow0^\circ$ is
singular as for $\theta > 0$ the Dirac point is fourfold degenerate,
while at $\theta=0$ one has the twofold
degeneracy of the $AB$ stacked bilayer. Interestingly, in this limit
the electronic properties
are in an essential way determined \emph{globally}, in contrast to the 
'nearsightedness'\cite{kohn96} of electronic structure generally found
in condensed matter.

\end{abstract}


\maketitle


\section{Introduction}

In addition to offering a possible route towards exploiting
the many remarkable properties of graphene\cite{geim07},
the epitaxial
growth of graphene on SiC\cite{berg06,ohta06,heer07,emt08}
presents a number of mysterious 
aspects. Principle amongst these is that the thermally 
induced growth of graphene on the C-face typically 
results in several graphene layers and yet, remarkably, this
complex graphene-based system shows behavior identical to
that of single layer graphene (SLG). In striking contrast,
bilayer graphene produced by mechanical exfoliation
has already a different low energy electronic
structure to that of SLG; a quadratic dispersion instead of linear.

An insight into this intriguing behavior of the C-face growth
was recently provided by Hass \emph{et al.}\cite{hass08}. These authors
showed that growth on the C face results in a high density
of twist boundary faults, i.e., layers with a relative rotation. 
Furthermore, \emph{ab-initio} calculations by the same authors
showed that if two graphene layers were rotated with the same relative
rotation observed in experiment, $\theta = 30^\circ \pm 2.20^\circ$,
then these layers exhibited a linear spectrum near the Dirac
point, exactly as in SLG. Rotation and translation
of graphene layers thus have profoundly different impact on the
low energy spectrum, and this lies at the heart of the C-face 
behavior.

While rationalizing the SLG nature of the C-face, these findings
raised a number of questions. Firstly, as to the character of the
rotational degree of freedom in few layer graphene systems: do
all rotations cause such an electronic decoupling or, alternatively,
only a subset of "magic" angles? This question is relevant to 
experiments as subsequent investigations have shown that various
angles of rotation may occur during growth on the C-face\cite{varch08,emt08}.
Clearly, a related question is the nature of the mechanism 
responsible for this electronic decoupling: how does the rotation
lead to the emergence of an effective Dirac-Weyl equation
for low energies?

These questions, at first sight, appear difficult from the
point of view of theory as one ultimately requires \emph{general} statements
to be made about an \emph{infinite} class of possible lattices.
Initially, theoretical progress was made by example of specific 
rotation angles or limits,
with graphene bilayer and trilayer systems calculated \emph{ab-initio}
in Ref.~[\onlinecite{lat07}], while in Ref.~[\onlinecite{san07}] 
the $\theta\rightarrow0^\circ$ limit of the twist bilayer was investigated via 
a continuum approximation to the tight binding Hamiltonian.
In the former case a low energy linear spectrum was noted
for all layers experiencing a relative rotation, while the latter work found also
a linear spectrum but with the Fermi velocity at the Dirac point,
$v_F$, strongly suppressed 
as compared to SLG. Subsequent Raman spectroscopy
experiments\cite{ni08,pon08} differ on whether this effect is present
in misoriented graphene samples; in Ref.~[\onlinecite{ni08}] a blueshift
of the graphene 2D peak was attributed to this effect, however 
in Ref.~[\onlinecite{pon08}] this was instead attributed to a modification
of the phonon dispersion in misoriented layers.

In Ref.~[\onlinecite{shall08}] it was shown
that the rotational degree of freedom was associated with a destructive 
interference of quantum states from each layer, and that this resulted
in a coupling that becomes progressively weaker as the size of the
commensuration cell increases. In fact, coupling at the Dirac point is
already very weak for the smallest possible commensuration,
a cell of 28 carbon atoms, with a splitting of $7\,$meV found
in \emph{ab-initio} calculations\cite{shall08}. All misoriented graphene layers are,
therefore, predicted to show effectively decoupled Dirac cones.

Further theoretical investigations have
been undertaken with regard to both the energetics
of misoriented layers\cite{shall08a}, and the
simulation of scanning tunneling microscopy images for
such layers\cite{cis08}. In the former work it was
noted that the sliding energy of relatively rotated
graphene layers is essentially zero, in dramatic contrast
to the case without rotation where the AB configuration
is energetically favored.
Most recently, tight-binding
calculations have been performed for a wide range
of misorientation angles\cite{tra09}. This latter
work demonstrates a reduction in the Fermi velocity that, for a wide
range of rotation angles, agrees with the result of
Ref.~[\onlinecite{san07}].

In this article we aim to accomplish two things. Firstly,
the formalism presented in Ref.~[\onlinecite{shall08}] is extended
to explain, on general lattice grounds (i.e., without deploying a 
continuum approximation), both the Dirac cone decoupling and
Fermi velocity suppression. Secondly, we provide a numerical
implementation of this formalism using the tight-binding
method. We demonstrate that this numerical scheme is at least an order of
magnitude faster than the usual tight-binding basis, and using
this explore the electronic structure as a function of
rotation angle for $1.47^\circ \le \theta < 30^\circ$.

We now present a brief summary of the content of this article.
Firstly, in Section II we discuss in detail the
crystal structure of mutually rotated graphene layers, and
derive the conditions for a commensurate crystal structure to occur.
An important feature of this system is the emergence, for $\theta \lesssim 15^\circ$,
of a so-called moir\'e pattern\cite{camp07}. This is a hexagonal interference
pattern, consisting of regions of AA and AB stacking, the periodicity
of which represents a new structural length scale of the system.

Section III then describes the electronic
structure of the bilayer in terms of a basis 
formed from the quantum states of the two mutually rotated layers
with, in addition, the
bilayer one-electron potential treated as a superposition
of two single layer potentials, i.e. $V^{(1)}+\bR V^{(2)}$ with
$\bR$ the rotation operator and $V^{(1,2)}$ potentials with
the in-plane translation symmetry of SLG, an approach
first described in Ref.~[\onlinecite{shall08}].
It is
shown how this leads to a convenient separating out of
purely symmetry related aspects of the electronic structure,
leading to simple conditions for determining if
the overlap elements of the potential with single layer states
are vanishing or not. As the interlayer interaction part
of the full bilayer Hamiltonian
may be constructed from such overlap elements,
an understanding of how and why these vanish leads in turn
to an understanding of the nature of the interlayer decoupling
in this system.

In this context we investigate how the overlap between states
from the constituent layers
depends on their $\bk$-vectors (i.e., their
$\bk$-vectors in the two mutually rotated
single layer Brillouin zones). We find that this
dependence is rather subtle, and that the vanishing
or not of such overlaps depends crucially on
these $\bk$-vectors. On this basis we demonstrate a number
of general features of the bilayer electronic structure,
and in particular show that
(i) for the Dirac bands the 1st order term of a
perturbation theory in the interlayer interaction is negligible
for all rotations and that, furthermore, 
(ii) 2nd order order terms in perturbation theory lead to 
a Fermi velocity suppression of the form found
in Ref.~[\onlinecite{san07}]. We further develop two consequences of
(i); if 2nd (and higher) order terms are unimportant then
the Dirac cones effectively decouple, and that the Dirac bands
from each layer are degenerate regardless of the role of
higher order terms.

Discussed also in this Section is the rather unusual
$\theta \rightarrow 0$ limit, which is a singular limit as
for any $\theta > 0$ the electronic structure
is dramatically different from that at $\theta = 0$. This is,
of course, simply a electronic manifestation of the
fact that the lattice geometry is also singular in this limit:
for any small but non-zero $\theta$ one has a moir\'e pattern,
while at $\theta = 0$ the graphene layers are
simply AB (or AA) stacked. An interesting aspect of this limit
is a breakdown in the notion of 'nearsightedness', i.e., that
electronic properties are essentially determined locally. As
the moir\'e periodicity diverges as $\theta\rightarrow0^\circ$ and, furthermore,
as the twist bilayer electronic structure must, for any finite $\theta$, 
be different from both the AA and AB stacked bilayers, 
one concludes that in this limit the electronic properties 
are, in contrast, in an essential way determined globally.

Finally, Section IV is devoted to a presentation of 
tight-binding calculations for the graphene
twist bilayer. We demonstrate that a basis
formed by the quantum states of the two mutually 
rotated layers converges remarkably
quickly, and leads to a dramatic improvement in
computational efficiency. Using this we then
investigate the bilayer electronic structure
for $1.47^\circ \le \theta < 30^\circ$ and
find a suppression of the Fermi velocity, $v_F$,
that is dramatic for small angles (at $\theta=1.47^\circ$ the reduction
in $v_F$ is 95\%) but, in agreement with all
\emph{ab-initio} calculations to date\cite{hass08,lat07,shall08},
insignificant for $\theta > \approx15^\circ$.
However, while the expression for the Fermi velocity suppression derived here and
in Ref.~[\onlinecite{san07}], describes
almost perfectly the tight-binding results for $5^\circ \lesssim \theta<30^\circ$,
it breaks down for $\theta \lesssim 5^\circ$. 
This breakdown is a result of the fact that the Fermi 
velocity suppression is a 2nd order effect in layer interaction, 
and in the limit $\theta\rightarrow0^\circ$ the resulting near degeneracy 
of the Dirac cones entails the importance of terms beyond this
order.


\section{Commensuration conditions of the twisted bilayer}


\begin{figure}
\caption{(Color online) Shown is the number of C atoms, $N_C$, in the commensuration cell
as a function of the relative orientation of the two graphene layers,
for $N_{C}<7000$. Inset displays the moir\'e pattern for
the cell indicated number 4. Band structures of twist bilayers
corresponding to the points labeled 1-4 are displayed in panels
1-4 of Fig.~\ref{fig:band}. The dashed line corresponds the lower bound
$N_{C}=(\sin^2\theta/2)^{-1}$; for commensuration cells
that fall on this line the moir\'e periodicity is equal to the
commensuration periodicity, see Section II for details.
\vspace{1.5cm}
}
{\includegraphics[scale=0.50,angle=00]{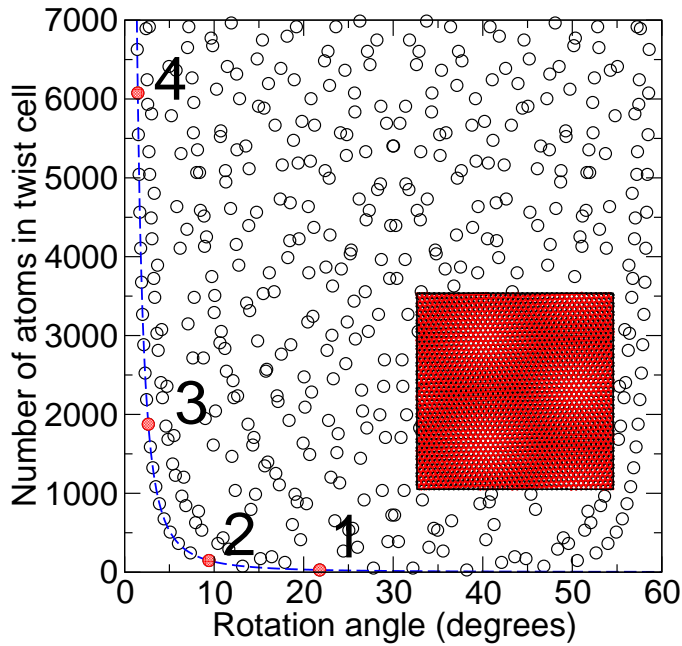}}
\label{fig:cslN}
\end{figure}

\begin{figure}
\caption{(Color online) Illustration of the commensuration cell for the case of
a misorientation angle of $\theta=21.78^\circ$, generated by a $(p,q)$ pair
of (1,3) (lattice vectors $\bt_1$ and $\bt_2$).
Shown also are the unit cells of the unrotated graphene layer
(vectors $\ba_1$, $\ba_2$) and rotated graphene layer (vectors
$\bR\ba_1$, $\bR\ba_2$). For explanations of other symbols refer
to Section 2.
\vspace{1.5cm}
}
{\includegraphics[scale=0.50,angle=00]{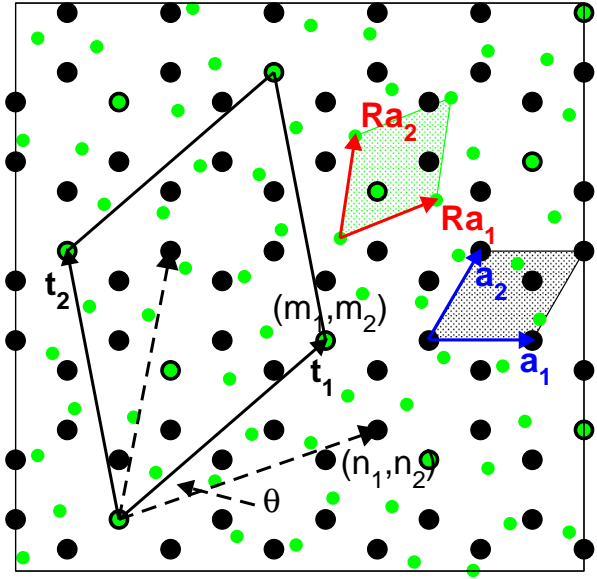}}
\label{fig:csl}
\end{figure}

A prerequisite to exploring the electronic structure of the
twisted bilayer is an elucidation of the crystallography
of such a system, i.e., determining the conditions under
which two misoriented layers are in commensuration. This problem
was studied in Ref.~[\onlinecite{shall08}] where a complete solution
was presented; here we provide a more detailed derivation of those
results with, additionally, a somewhat simpler and more symmetrical
choice of commensuration vectors.

Evidently, the \emph{existence} of a commensuration depends only on the 
relative rotation of the lattice vectors
of each layer, and not on the structure of the unit cells of each layer.
Thus we need not, at this stage, concern ourselves with which axis the rotation
is taken
about and the initial configuration (AB or AA, and so on) of the graphene bilayer;
these amount to different choices of initial basis vectors within each cell. The
commensuration condition may be written as

\begin{equation} \label{eq:rc}
\br_1 = \bR \br_2
\end{equation}
where $\br_1,\br_2$ are hexagonal lattice vectors, and $\bR$ the rotation
operator. The set $\{\br_1\}$ are the resulting coincident points between
the two layers, while $\{\br_2\}$ is the same set,
but viewed from the rotated coordinate system. Utilizing 
the unrotated lattice as a coordinate system, i.e.,
$\br = i \ba_1 + j \ba_2$ with $i,j\in\mathbb{Z}$, Eq.~(\ref{eq:rc}),
with a standard choice of lattice vectors
$\ba_1=[\sqrt{3},0]$ and $\ba_2 = [\frac{\sqrt{3}}{2}, 3/2]$, 
may be written as

\begin{equation} \label{eq:dio1}
\begin{pmatrix} m_1 \\ m_2 \end{pmatrix} = 
\begin{pmatrix}
\cos\theta - \frac{1}{\sqrt{3}} \sin\theta & -\frac{2}{\sqrt{3}} \sin\theta \\
\frac{2}{\sqrt{3}} \sin\theta & \cos\theta +\frac{1}{\sqrt{3}}\sin\theta
\end{pmatrix}
\begin{pmatrix} n_1 \\ n_2 \end{pmatrix}.
\end{equation}
(Here and throughout this article our unit of length is chosen to be the 
graphene C-C separation.)
This maps one integer pair $(n_1,n_2)$ to another $(m_1,m_2)$
and, for this to be possible, a necessary and sufficient condition on the 
matrix in Eq.~(\ref{eq:dio1}) is that it assumes only rational values\cite{for83}.
This leads to the
following conditions on $\theta$

\begin{eqnarray}
\frac{1}{\sqrt{3}} \sin\theta & = & \frac{i_1}{i_3} \label{eq:DD1} \\
\cos\theta & = & \frac{i_2}{i_3}\label{eq:DD2},
\end{eqnarray}
where $i_1,i_2,i_3\in\mathbb{N}$. These are therefore related
by the following second order Diophantine equation

\begin{equation} \label{eq:dio2}
3i_1^2 + i_2^2 = i_3^2.
\end{equation}
Solution of
this equation proceeds in a standard way\cite{ste05} (analogous to the
case of Pythagorean triples) by dividing by $i_3^2$ and making
the substitution $x = \frac{i_1}{i_3}$, $y = \frac{i_2}{i_3}$.
There is thus a one to one mapping between solutions of Eq.~(\ref{eq:dio2})
and rational points on the ellipse $3x^2+y^2=1$. One such point is
$(0,1)$ and any other may be found by the intersection with the ellipse 
of a line passing through $(0,1)$ and $(q/p,0)$.
The coordinates of this latter point then lead to the following
solution for $i_1,i_2,i_3$

\begin{eqnarray}
i_1 & = & 2pq \label{eq:dio3a} \\
i_2 & = & 3q^2-p^2 \label{eq:dio3b} \\
i_3 & = & 3q^2+p^2 \label{eq:dio3c}
\end{eqnarray}
where $p,q\in\mathbb{N}$.
From these equations we immediately
find the set of rotation angles leading to commensurations,

\begin{equation} \label{eq:dioth}
\theta = \cos^{-1}\left(\frac{3q^2-p^2}{3q^2+p^2}\right).
\end{equation}
For $q \ge p > 0$ this
formula produces rotation angles that lie in the range
$0 \le \theta \le 60^\circ$. All other rotation angles are equivalent due to the symmetry of the 
hexagonal lattice. Clearly, the limit $p/q \rightarrow 0$
corresponds to $\theta\rightarrow 0^\circ$ while, on the other hand,
the limit $p/q\rightarrow 1$ corresponds to $\theta\rightarrow 60^\circ$.
Note that changing the sign of $p$ or $q$ sends $\theta\rightarrow-\theta$.
Since the limit $\theta\rightarrow60^\circ$ is equivalent to
$\theta\rightarrow0^\circ$ taken from below, and since
formally little changes by the substitution $p\rightarrow-p$ or
$q\rightarrow-q$, our focus in this work will be on
angles in the range $0^\circ \le \theta \le 30^\circ$.

We also require the corresponding primitive vectors of the commensuration lattice.
Substitution of Eqs.~(\ref{eq:DD1}) and (\ref{eq:DD2}) into Eq.~(\ref{eq:dio1})
results in the following coupled linear Diophantine equations

\begin{eqnarray} \label{eq:dio4}
\begin{pmatrix} m_1 \\ m_2 \end{pmatrix} = \frac{1}{i_3} 
\begin{pmatrix} i_2-i_1 & -2i_1 \\
                2i_1 & i_2+i_1 \end{pmatrix}
\begin{pmatrix} n_1 \\ n_2 \end{pmatrix}.
\end{eqnarray}
The solution of these equations follows by a similarity transform such that
the matrix multiplying $(n_1 n_2)^{T}$ is diagonal. Crucially, the eigenvectors
of this matrix are \emph{independent} of $p,q$ and thus the problem is recast
as coupled linear diophantine equations that are \emph{linear} in $p,q$. These
may then be solved by inspection yielding the result that

\begin{eqnarray}
\begin{pmatrix} n_1 \\ n_2 \end{pmatrix}
& = & \alpha \begin{pmatrix}  p+3q \\  -2p \end{pmatrix}
 + \beta \begin{pmatrix} 2p \\  -p+3q \end{pmatrix} 
\label{eq:dio6a} \\
\begin{pmatrix} m_1 \\ m_2 \end{pmatrix}
& = & \alpha \begin{pmatrix}  -p+3q \\ 2p \end{pmatrix}
 + \beta \begin{pmatrix}  -2p \\  p+3q \end{pmatrix}
\label{eq:dio6b},
\end{eqnarray}
with $\alpha,\beta$ are arbitrary constants such that $\bnn$ and $\bem$
are integer valued. The final step is to determine the primitive
vectors of the commensuration lattice. This calculation
we present in Appendix A, and here quote only the result.
The form of the commensuration vectors turns out to depend on a parameter
$\delta = 3/\GCD(p,3)$. For the case where $\delta = 1$ we find

\begin{equation} \label{eq:pv1}
\bt_1 = \frac{1}{\gamma} \begin{pmatrix} p + 3q \\ - 2p \end{pmatrix},
\bt_2 = \frac{1}{\gamma} \begin{pmatrix} 2p \\ -p+3q \end{pmatrix}
\end{equation}
while for the case $\delta=3$ we find

\begin{equation} \label{eq:pv3}
\bt_1 = \frac{1}{\gamma} \begin{pmatrix} -p-q \\ 2q \end{pmatrix},
\bt_2 = \frac{1}{\gamma} \begin{pmatrix} 2q \\ -p+q \end{pmatrix},
\end{equation}
where $\gamma=\GCD(3q+p,3q-p)$. Values that this parameter may take
when $\GCD(p,q)=1$ are indicated in Table \ref{tab:gam}.

\begin{table}
\caption{\label{tab:gam} Possible values that the parameter $\gamma$
can take.
}
\begin{minipage}{0.5\textwidth}
\begin{ruledtabular}
\begin{tabular}{l|cc}
              & $\delta = 1$ & $\delta = 3$ \\ \hline
p, q odd      &  6           &   2          \\
otherwise     &  3           &   1
\end{tabular}
\end{ruledtabular}
\end{minipage}
\end{table}

Thus every possible commensuration between misoriented layers
is \emph{uniquely specified} by an integer pair $p \ge q > 0$ such that $\GCD(p,q)=1$.
Given this we can completely characterize
the commensuration; the rotation angle may be obtained
from Eq.~(\ref{eq:dioth}), while the lattice vectors are given by
either Eq.~(\ref{eq:pv1}) or
Eq.~(\ref{eq:pv3}),
depending on whether the parameter $\delta=3/\GCD(p,3)$
assumes the values of 1 or 3 respectively. The
various notations introduced in this derivation are
illustrated in Fig.~\ref{fig:csl}.

It is worth reflecting on the reason that two integers, $p$ and $q$,
are needed to specify a commensuration while, on the other hand
it is clear that any bilayer lattice (commensurate or incommensurate)
is uniquely specified by a single number $\theta$.
This is a consequence of the relation between the real and rational number
fields: given a $\theta$ there are infinitely many choices of $p$ and $q$
in Eq.~\ref{eq:dioth} such that $\theta$ may be reproduced to an
arbitrary accuracy $\epsilon$.

The ratio of the total number of lattice vectors to 
coincident lattice vectors for the twist boundary
is found to be given by

\begin{equation} \label{eq:dioN}
N = \frac{|(\bt_1\times\bt_2).\hat{\bz}|}{|(\ba_1\times\ba_2).\hat{\bz}|}
 = \frac{3}{\delta}\frac{1}{\gamma^2} (3q^2+p^2),
\end{equation}
with the number of carbon atoms in the commensuration cell
$N_{C}=4N$.
(The factor 4 simply arising from the fact there are two layers in the cell, 
and two basis atoms in the honeycomb structure.)
In Fig.~\ref{fig:cslN} is plotted $N_{C}$, 
as a function of misorientation angle;
the minimum $N_{C}$ is 28 corresponding to $\theta=30^\circ\pm8.21$,
however $N_{C}$ diverges in the $\theta\rightarrow0^\circ$ 
(or $\theta\rightarrow60^\circ$) limits.
Combining Eqs.~(\ref{eq:dioth}) and (\ref{eq:dioN}) we find that

\begin{equation} \label{eq:cslNth}
N = \frac{3}{\gamma^2\delta}\frac{p^2}{\sin^2\theta/2}.
\end{equation}
and so for $\theta\rightarrow0^\circ$ $N$ diverges as $1/\theta^2$.
This small angle limit is associated with the emergence of a new structural length
scale, that of the moir\'e periodicity $D$; such a moir\'e pattern
is shown in the inset of Fig.~\ref{fig:cslN}. The relation
between $D$ and $\theta$ is given by\cite{bey99}

\begin{equation} \label{eq:MM}
D=\frac{a}{2\sin{\theta/2}}.
\end{equation}
where $a$ is the graphene lattice constant.
The relation between the lattice constant of the 
commensuration cell and the moir\'e periodicity
may be seen by setting $p=1$, $\delta=3$, $\gamma=2$ in Eq.~(\ref{eq:cslNth}),
(corresponding to cells generated by $p=1$ and $q$ an 
odd integer, see Table \ref{tab:gam}),
and using $N=D^2/a^2$ which then gives back the formula for the
moir\'e periodicity, Eq.~(\ref{eq:MM}). In this case, therefore,
the moir\'e periodicity is equal to the commensuration cell
lattice constant. For these 
commensuration cells we find $N_{C}=(\sin^2\theta/2)^{-1}$,
and this is the lower bound function plotted in Fig.~\ref{fig:cslN}.
On the other hand,
for all other commensuration cells the "commensuration periodicity"
is greater than the moir\'e periodicity.

Finally, we note that the analytic results presented here are in agreement with the
numerical solution to this problem provided recently
by Campenara \emph{et al.}\cite{camp07}; special cases of
these results have been found in Ref.~[\onlinecite{san07}] (the
case $p=1$, $\delta=3$, $\gamma=2$) and more recently in Ref.~[\onlinecite{tra09}]
(the case $\delta=1$).


\section{Analysis of the interlayer interaction}

\begin{figure}
\caption{(Color online) Brillouin zones of the unrotated (U) and
rotated (R) graphene layers, as well as the Brillouin zone of the
bilayer supercell (B) for the case $(p,q)=(1,5)$, corresponding to
$\theta=13.17^\circ$. In this Figure
$\bb_1$, $\bb_2$ are the reciprocal lattice vectors of the unrotated
graphene layer, $\bR \bb_1$, $\bR \bb_2$ the  reciprocal lattice vectors of
the rotated layer, and $\bg_1$, $\bg_2$ the reciprocal lattice
vectors of the bilayer supercell. Special $K$ points
of various Brillouin zones indicated by subscript U, R, and B.
The separations of special $K$ points indicated are 
$\Delta K = |\bK_R-\bK_U|=|\bK_R^{*}-\bK_U^{*}|$,
$\Delta K_1 = |\bK_U^{*}-\bK_R|$, and $\Delta K_2 = |\bK_U-\bK_R^{*}|$.
\vspace{1.5cm}
}
{\includegraphics[scale=0.50,angle=00]{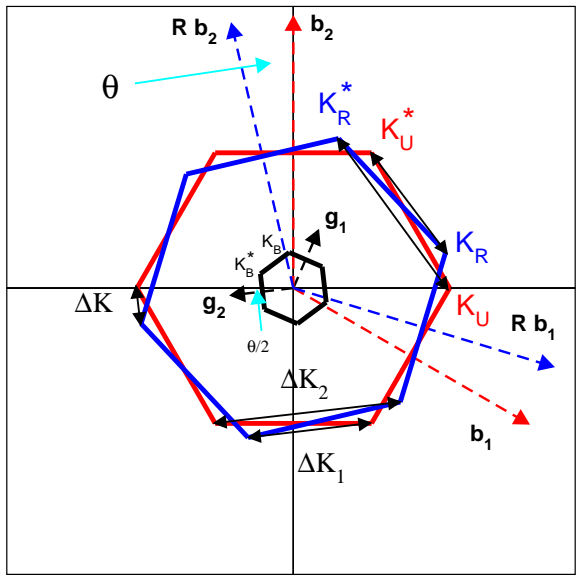}}
\label{fig:diobz}
\end{figure}

In this section we shall describe how the problem of understanding 
the \emph{general} electronic properties of the twist bilayer
\emph{for any $\theta$} is solved. Our approach is that
described in Ref.~[\onlinecite{shall08}] which, in broad outline,
may be characterized as 'constructing the bilayer system from single layer
components'. We take the bilayer potential as a superposition
of single layer potentials, and use as a basis for this new system the eigenkets
of the single layer systems. The advantage of this is that the resulting
matrix elements may then be analyzed as a commensuration problem of
reciprocal space lattices. Such commensuration problems can be readily understood
for any angle, and thus one may then understand
the physics of the twist bilayer for general angle.

The remainder of this Section proceeds as follows. In Section III A we first
set up the Hamiltonian and basis used to analyze the twist bilayer.
Following this, in Section III B the
various reciprocal lattices and associated Brillouin zones
are described. In Section III C a condition is derived that
determines the vanishing of overlap elements found in the
model of Section III A.
Finally, in Sections III D-F, we use this understanding
to determine a number of generic electronic properties of the twist bilayer.

\subsection{Model Hamiltonian and basis}

In considering the 
interaction between misoriented layers
it is useful to take the bilayer potential as simply the superposition
of two single layer potentials, i.e., 

\begin{equation} \label{eq:Hsup}
\bH = \frac{\bpp^2}{2m} + \bV^{(1)} + \bV^{(2)}.
\end{equation}
Here $\bV^{(n)}$ are one-electron single layer graphene (SLG) potential operators that satisfy

\begin{equation} \label{eq:slh}
\bH^{(n)}\kh{i}{n} = (\frac{\bpp^2}{2m} + \bV^{(n)})\kh{i}{n}
= \ep{i}{n} \kh{i}{n}
\end{equation}
where $\ep{i}{n}$ are SLG eigenvalues, and $i$ and $\bk$ represent band 
and $\bk$-vector quantum numbers respectively.
These one-electron SLG potentials are invariant under different
in-plane translations;
we have $\bT \bV^{(1)} = \bV^{(1)}$ and $\bR\bT\bR^{-1} \bV^{(2)} = \bV^{(2)}$.
We shall take
the superscript "1" to denote objects associated with the unrotated
layer, and superscript "2" for objects associated with the rotated layer.
(Such a designation is, in \emph{mutually} rotated layers, clearly arbitrary and
made only for convenience.)
Given the weak interaction (and thus large separation) of the 
graphene layers, this approximation of the bilayer potential 
as a superposition of single layer potentials is expected to be good,
but, in any case, is certainly sufficient for the qualitative understanding 
presented here.

As a basis for this Hamiltonian we take the combined eigenkets 
of the unrotated and rotated layers, i.e.  
$\{\kn{1}\},\{\kn{2}\}$. One should note that since each 
SLG basis set by itself is complete on $\mathbb{R}^3$, this is generally 
an over-complete basis set.  On the other hand for minimal basis methods, such 
as the $p_z$ tight binding method in which the basis consists of a $p_z$
atomic orbital centered at every site in the crystal, a bilayer basis
set consisting of the combined eigenkets from each layer is 
clearly isomorphic to the usual basis set that would be employed.

\subsection{Reciprocal space properties of the bilayer system}

Here we describe the reciprocal space lattices corresponding to
the various real space lattices introduced in Section II. First
a note of nomenclature; we denote the reciprocal lattice
vectors corresponding to the unrotated (rotated) real space vectors
$\ba_1$ and $\ba_2$ ($\bR \ba_1$ and $\bR \ba_2$) by $\bb_1$ and $\bb_2$ 
($\bR \bb_1$ and $\bR \bb_2$), while the reciprocal space
lattice vectors corresponding to the the real space commensuration 
vectors $\bt_1$ and $\bt_2$ are denoted by $\bg_1$ and $\bg_2$. We shall
refer to this latter reciprocal space lattice as the
bilayer reciprocal lattice.

The vectors $\bg_1$ and $\bg_2$ are found from Eq.~(\ref{eq:pv1}) to be

\begin{eqnarray}
\bg_1 & = & \frac{\gamma}{3(3q^2+p^2)}[(p+3q)\bb_1 + 2p\bb_2] \label{eq:g1a} \\
\bg_2 & = & \frac{\gamma}{3(3q^2+p^2)}[-2p\bb_1 - (p-3q)\bb_2] \label{eq:g1b}
\end{eqnarray}
for the case where $\delta=1$ and
from Eq.~(\ref{eq:pv3}), to be

\begin{eqnarray}
\bg_1 & = & \frac{\gamma}{3q^2+p^2}[-(p-q)\bb_1 + 2q\bb_2] \label{eq:g3a} \\
\bg_2 & = & \frac{\gamma}{3q^2+p^2}[-2q\bb_1 - (p+q)\bb_2] \label{eq:g3b}
\end{eqnarray}
for the case $\delta=3$. The Brillouin zones associated with
each of these sets of primitive vectors, $\{\bb_1,\bb_2\}$, 
$\{\bR\bb_1,\bR\bb_2\}$, and $\{\bg_1,\bg_2\}$, are shown
in Fig.~\ref{fig:diobz} for the twist bilayer $(p,q)=(1,5)$. 
For convenience of exposition these Brillouin zones (BZ) will
be referred to by the abbreviations UBZ (for the unrotated BZ),
RBZ (for the rotated BZ), and BBZ for the BZ of the bilayer
reciprocal lattice.

These bilayer reciprocal lattice vectors determine a map
by which $\bk$-vectors in the UBZ and RBZ are related to those
of the BBZ (the usual so-called "folding back" condition of $\bk$-vectors).
It should be emphasized at this point that there are three separate 
$\bk$-indices in the problem as it is formulated here. 
We have a $\bk$-vector in the BBZ
which is a good quantum number for the bilayer Hamiltonian
and eigenkets, but we also have the $\bk$-indices of the single layer basis
used to solve the Hamiltonian at this $\bk$,
labeled by $\bk_1$ and $\bk_2$. To solve the Hamiltonian
at $\bk$, the single layer basis then consists of all those
eigenkets which map back from the UBZ and RBZ to the point $\bk$ in the BBZ.

An interesting, and for the nature of the interlayer interaction crucial,
relationship exists between the special K-points of all these BZ's:
to each special K-point of the BBZ is mapped back one of the
special K-point from the UBZ and one from the RBZ. The precise manner
in which this happens depends in a rather complex way
on the $p$, $q$ parameters of the commensuration, detailed in 
Tables \ref{tab:kmap1} and \ref{tab:kmap3}. The overall scheme, however, is clear:
for commensurations characterized by $\delta=1$ $K$-points connected by $\Delta K$ 
(see Fig.~\ref{fig:diobz}) map to the same special K-point of the BBZ while,
for $\delta=3$, it is the conjugate pairs $(K_U^{*},K_R)$ and
$(K_U,K_R^{*})$, separated by $\Delta K_1$ and $\Delta K_2$ respectively, 
that map back to the same special K-points of the BBZ
(these special K-points and their separations are 
indicated also in Fig.~\ref{fig:diobz}).

\begin{table}
\caption{\label{tab:kmap1} Structure of the mapping of special K points of the unrotated (U) and
rotated (R) layer Brillouin zones to the special K points of the bilayer (B) Brillouin zone,
for the case $\delta=1$. The designation of the special K-points corresponds
to that of Fig.~\ref{fig:diobz}.
}
\begin{ruledtabular}
\begin{tabular}{lcc} 
                 & $\gamma=3$                         & $\gamma=6$                         \\ \hline
$\mod(q,3)=1$    & $\bK_U     \rightarrow \bK_B     $ & $\bK_U     \rightarrow \bK_B^{*} $ \\
                 & $\bK_U^{*} \rightarrow \bK_B^{*} $ & $\bK_U^{*} \rightarrow \bK_B     $ \\
                 & $\bK_R     \rightarrow \bK_B     $ & $\bK_R     \rightarrow \bK_B^{*} $ \\
                 & $\bK_R^{*} \rightarrow \bK_B^{*} $ & $\bK_R^{*} \rightarrow \bK_B     $ \\ \hline

$\mod(q,3)=2$    & $\bK_U     \rightarrow \bK_B^{*} $ & $\bK_U     \rightarrow \bK_B     $ \\
                 & $\bK_U^{*} \rightarrow \bK_B     $ & $\bK_U^{*} \rightarrow \bK_B^{*} $ \\
                 & $\bK_R     \rightarrow \bK_B^{*} $ & $\bK_R     \rightarrow \bK_B     $ \\
                 & $\bK_R^{*} \rightarrow \bK_B     $ & $\bK_R^{*} \rightarrow \bK_B^{*} $ \\ \hline
\end{tabular}
\end{ruledtabular}
\end{table}

\begin{table}
\caption{\label{tab:kmap3} Structure of the mapping of special K points of the unrotated (U) and
rotated (R) layer Brillouin zones to the special K points of the bilayer (B) Brillouin zone,
for the case $\delta=3$. The designation of the special K-points corresponds
to that of Fig.~\ref{fig:diobz}.
}
\begin{ruledtabular}
\begin{tabular}{lcc} 
                 & $\gamma=1$                         & $\gamma=2$                         \\ \hline
$\mod(p,3)=1$    & $\bK_U     \rightarrow \bK_B^{*} $ & $\bK_U     \rightarrow \bK_B     $ \\
                 & $\bK_U^{*} \rightarrow \bK_B     $ & $\bK_U^{*} \rightarrow \bK_B^{*} $ \\
                 & $\bK_R     \rightarrow \bK_B     $ & $\bK_R     \rightarrow \bK_B^{*} $ \\
                 & $\bK_R^{*} \rightarrow \bK_B^{*} $ & $\bK_R^{*} \rightarrow \bK_B     $ \\ \hline

$\mod(p,3)=2$    & $\bK_U     \rightarrow \bK_B     $ & $\bK_U     \rightarrow \bK_B^{*} $ \\
                 & $\bK_U^{*} \rightarrow \bK_B^{*} $ & $\bK_U^{*} \rightarrow \bK_B     $ \\
                 & $\bK_R     \rightarrow \bK_B^{*} $ & $\bK_R     \rightarrow \bK_B     $ \\
                 & $\bK_R^{*} \rightarrow \bK_B     $ & $\bK_R^{*} \rightarrow \bK_B^{*} $ \\ \hline
\end{tabular}
\end{ruledtabular}
\end{table}

Thus without any layer interaction, i.e. what translational
symmetry alone requires, is that the Dirac cones from the
unrotated and rotated layers are mapped to the special
K-points of the BBZ. With no layer interaction we therefore find
two degenerate Dirac cones situated at each special K-point
of the BBZ. It should be stressed that this mapping
is particular to the $K$ star; a similar map
does not, for example, exist for the $M$ star.
Interestingly, this implies that $M$-point chiral fermions
(which may be generated by the application of 
a periodic scalar potential\cite{park08a}) would behave rather
differently in this context.

We may now consider what happens to this degeneracy
when we turn on a layer interaction. In general, of course,
such an interaction would result in a splitting of the Dirac cones,
however this is not what happens for the case of mutually
rotated graphene layers. The key to understanding this,
as we now describe,
lies in the remarkable behavior of the overlap elements
of the bilayer Hamiltonian, Eq.~(\ref{eq:Hsup}), with
states from the mutually rotated graphene layers.


\subsection{Matrix elements of the bilayer Hamiltonian}
\label{sec:mat}

Given a bilayer potential of the form $\bV^{(1)}+\bV^{(2)}$,
and a basis set of single layer eigenkets
$\{\kn{1}\},\{\kn{2}\}$, the electronic structure will be 
determined by \emph{interlayer}
matrix elements of the type $\bn{1} \bV^{(1)} \kn{2}$.
Using this matrix element as a specific example, we now show
how one may derive a general condition that determines whether
such a matrix element vanishes or not. Using
a plane wave expansion for each of the objects in this matrix element,
i.e.

\begin{eqnarray}
\bV^{(1)} & = & \sum_{\bG_1^{'}} V_{\bG_1^{'}}^{(1)} e^{i\bG_1^{'}.\br} \\
\phi_{i_1\bk_1}^{(1)\ast}(z) & = & \sum_{\bG_1^{''}} c^{(1)\ast}_{i_1 \bk_1+\bG_1^{''}}(z) e^{-i(\bk_1+\bG_1^{''}).\br} \\
\phi_{i_2\bk_2}^{(2)}(z) & = &\sum_{\bR\bG_2} c^{(2)}_{i_2 \bk_2+\bR\bG_2}(z) e^{i(\bk_2+\bR\bG_2).\br}
\end{eqnarray}
we find

\begin{eqnarray}
\bn{1} \bV^{(1)} \kn{2} & = &
\sum_{\bG_1,\bR\bG_2} \left( \sum_{\bG_1^{'}} \int\!dz\; 
 c^{(1)\ast}_{i_1 \bk_1+\bG_1+\bG_1^{'}}(z) V_{\bG_1^{'}}^{(1)}(z) c^{(2)}_{i_2 \bk_2+\bR\bG_2}(z)
 \right) \delta_{\bk_1+\bG_1 = \bk_2 + \bR\bG_2} \nonumber \\
& = & \sum_{\bG_1,\bR\bG_2} C_{\bk_1+\bG_1}^{\bk_2+\bR\bG_2}
 \delta_{\bk_1+\bG_1 = \bk_2 + \bR\bG_2}
\label{eq:msum}
\end{eqnarray}
where we have made the convenient substitution $\bG_1^{''}-\bG_1^{'}=\bG_1$. 
As the structure of Eq.~(\ref{eq:msum}) arises simply from the differing 
in-plane translation groups of the constituent objects, any such interlayer matrix
element may be cast into this form (although the 
coefficients $C_{\bk_1+\bG_1}^{\bk_2+\bR\bG_2}$ will obviously be different).

Clearly, it is the Kronecker delta term in Eq.~(\ref{eq:msum}) that is the most significant
consequence of rotation.
This Kronecker delta term ensures that in the double sum over
$\bG_1$ and $\bR\bG_2$ only terms satisfying 

\begin{equation} \label{eq:comcond}
\bG_1 = \bR\bG_2 + \bk_2-\bk_1
\end{equation}
contribute. This is just a commensuration condition between
the reciprocal lattices of the unrotated and rotated layers. However,
in contrast to the real space commensuration condition, $\ba_1 = \bR\ba_2$,
this involves not only the geometry via the $\bR$ operator, but also
a dependence on the single layer states through the term $\bk_2-\bk_1$.
This removal of
contributions from the interlayer matrix elements is a direct consequence of
the mutual rotation of the layers, and can be seen as a destructive
interference of the quantum states from each layer.
It is now clear that the advantage of the approach deployed here
is that we have separated the symmetry aspects of the problem,
which generate a \emph{selection condition} for the
coefficients $C_{\bk_1+\bG_1}^{\bk_2+\bR\bG_2}$,
from details of the electronic structure which are contained
in the actual values of these coefficients. 

Continuity of wavefunctions and potentials in real space
implies that these coefficients will decay to zero with increasing $|\bG_{1,2}|$
and will be largest for $\bG_{1,2}$ at or near
the origin. In addition, the coincident points between the lattices
$\bG_1$ and $\bR\bG_2+\bk_2-\bk_1$ become increasingly separated as $\theta\rightarrow0^\circ$
(just as the size of the real space commensuration cell diverges in this limit,
see Section II). Since
these coincident points represent the only symmetry allowed contributions
to the bilayer matrix elements, we see that for sufficiently
small $\theta$ there can be \emph{at most one contributing
term}, which occurs in the case of a coincident point close 
to the origin of reciprocal space. For example, if
$\bk_1=\bk_2$ in Eq.~(\ref{eq:comcond}) then $\bG_1=\bR\bG_2=0$
is a solution and, as the coefficient
$C_{{\bf 0}}^{{\bf0}}$ is generally non-zero, 
so in turn will the matrix element will
remain finite for all $\theta$. On the other hand, in the
case where all coincident points
are sufficiently far from the origin, the matrix element
will vanish for sufficiently small $\theta$.
(An illustration of these two cases is given in the left and right
hand side panels of Fig.~\ref{fig:kshfts}).

As we shall now show the term
$\bk_2-\bk_1$ in Eq.~(\ref{eq:comcond}) results
simply in a shift from the origin of the commensuration lattice (i.e., the lattice
of coincident points) that would be found for the
case $\bk_2-\bk_1={\bf 0}$. The relation between
the term $\bk_2-\bk_1$ and this shift, which can be found
by solving Eq.~(\ref{eq:comcond}), thus plays a crucial role
in determining which bilayer matrix elements will vanish.

The solution follows by the casting of Eq.~(\ref{eq:comcond})
into a Diophantine problem, exactly as
outlined in Section II for the real space case. This, and
the solution of the resulting Diophantine problem,
are described in Appendix B. Here we summarize the results
with the solutions
expressed in terms of the unrotated reciprocal
lattice vectors as $m_1 \bb_1 + m_2 \bb_2$.
One finds that for the case $\delta = 1$ two possible solutions given by

\begin{equation} \label{eq:msol1}
\bem = \alpha \frac{1}{\gamma} \begin{pmatrix} p + 3q \\ 2p \end{pmatrix}
 + \beta \frac{1}{\gamma} \begin{pmatrix} -2p \\ -p+3q \end{pmatrix}
 + \frac{\gamma}{6q} \begin{pmatrix} l_1 \\ l_2 \end{pmatrix}.
\end{equation}
and

\begin{equation} \label{eq:msol2}
\bem = \alpha \frac{1}{\gamma} \begin{pmatrix} p + 3q \\ 2p \end{pmatrix}
 + \beta \frac{1}{\gamma} \begin{pmatrix} -2p \\ -p+3q \end{pmatrix}
 + \frac{\gamma}{6p} \begin{pmatrix} l_1-2l_2 \\ 2l_1-l_2 \end{pmatrix}.
\end{equation}
While for the case $\delta = 3$ one finds

\begin{equation} \label{eq:msol3}
\bem = \alpha \frac{1}{\gamma} \begin{pmatrix} -p+q \\ 2q \end{pmatrix}
 + \beta \frac{1}{\gamma} \begin{pmatrix} 2q \\ p+q \end{pmatrix}
 - \frac{\gamma}{2p} \begin{pmatrix} l_1 \\ l_2 \end{pmatrix}
\end{equation}
and

\begin{equation} \label{eq:msol4}
\bem = \alpha \frac{1}{\gamma} \begin{pmatrix} -p+q \\ 2q \end{pmatrix}
 + \beta \frac{1}{\gamma} \begin{pmatrix} 2q \\ p+q \end{pmatrix}
 + \frac{\gamma}{6q} \begin{pmatrix} l_1-2l_2 \\ 2l_1-l_2 \end{pmatrix}
\end{equation}
Here $\alpha$ and $\beta$ are arbitrary integers, $\gamma=\GCD(3q+q,3q-p)$, and $l_1$
and $l_2$ are the integers that result when $\bk_2-\bk_1$ is
expressed in coordinates of the bilayer reciprocal lattice, i.e.

\begin{equation} \label{eq:l1l2}
\bk_2-\bk_1 = l_1\bg_1+l_2\bg_2. 
\end{equation}
(Note that since both $\bk_1$ and $\bk_2$ fold back,
under translations by the bilayer reciprocal lattice vectors $\bg_1$ and $\bg_2$,
to the same $\bk$-point of the BBZ then their difference
can be expressed as integer multiples of $\bg_1$ and $\bg_2$.
Hence in coordinates of the bilayer reciprocal lattice
the difference $\bk_2-\bk_1$ will always be integer.)
Clearly, in all cases the solutions are of the form

\begin{equation}
\alpha\tilde{\bG}_1 + \beta\tilde{\bG}_2 + \Delta\tilde{\bG},
\end{equation}
with the important constant shift $\Delta\tilde{\bG}$ determined
by $\bk_2-\bk_1$.

It should be noted that these expressions 
provide only a partial solution to 
Eq.~(\ref{eq:comcond}). The reason is for this is that while $(m_1,m_2)$ must
be integer valued, the shift terms are obviously not integer
valued unless, e.g., both $l_1$ and $l_2$ are divisible by
$6q/\gamma$ in Eq.~(\ref{eq:msol1}). This absence of a complete
solution is due to the fact that,
as shown in Appendix B, Eq.~(\ref{eq:comcond}) results in
an \emph{inhomogeneous} simultaneous linear Diophantine problem (in contrast 
to the homogeneous problem of the real space commensuration) which is 
known to have no analytic solution. However, as we now demonstrate,
this partial solution provides sufficient insight into the selection rule, 
Eq.~(\ref{eq:comcond}), that several generic features of the bilayer 
electronic structure may be elucidated.


\subsection{Decoupling of the Dirac cones}
\label{sec:decup}

Here we shall prove that for all commensuration cells
with $N_{C}$ \emph{greater than some critical value} the
Dirac bands from the unrotated and rotated layers will
be effectively degenerate in energy. Thus there will be a fourfold
degeneracy at the Dirac points of the bilayer band structure, with the Dirac
bands themselves twofold degenerate. While this derivation demonstrates
the \emph{existence} of such a critical value, the numerical value of
this parameter will depend on the coefficients $C_{\bk_1+\bG_1}^{\bk_2+\bR\bG_2}$ and can,
of course, only be determined by actual calculation of the
electronic structure. 

It should be stressed that this result requires
only (i) mutually rotated and weakly interacting 
layers of hexagonal symmetry and (ii) the low energy spectrum 
to located at the vectors of the K star; it is, therefore,
applicable to other contexts in which graphene may be created.

Our approach is based on a perturbative treatment and the use of the
selection rules for terms in the matrix element sums derived in the
previous section. Since in the absence of
any interlayer interaction we have two degenerate Dirac cones at the
special $K$-points of the BBZ (see Section III B), then the first order energy
shift will
be given by the secular equation of degenerate state perturbation
theory. 

One should note that, without interaction, the degeneracy at and away from the
Dirac points will be different: fourfold at the Dirac point and twofold away. 
In fact, as our considerations are entirely based
on the in-plane translation groups, and not on point group
symmetry arguments, then there will be no fundamental difference
in how we treat these two cases, and for simplicity we consider here
the case of a twofold degeneracy. The resulting secular equation
is then

\begin{equation} \label{eq:seceq}
\begin{pmatrix}
\delta \tH_{11} & \delta \tH_{12} \\
\delta \tH_{21} & \delta \tH_{22}
\end{pmatrix}
\begin{pmatrix} a_1 \\ a_2 \end{pmatrix} =
\delta\epsilon^{(1)}_i
\begin{pmatrix} a_1 \\ a_2 \end{pmatrix}
\end{equation}
where the elements $\delta \text{H}_{ij}$ are given by

\begin{eqnarray}
\delta \tH_{11} & = & \bn{1} \bV^{(2)} \kn{1} \label{eq:mat1} \\
\delta \tH_{12} & = & \frac{1}{2}(\epsilon^{(1)}_{i_1\bk_1} + \epsilon^{(2)}_{i_2\bk_2}) \prdn{1}{2} + \bn{1} \overline\bV \kn{2} \label{eq:mat2} \\
\delta \tH_{21} & = & \frac{1}{2}(\epsilon^{(1)}_{i_1\bk_1} + \epsilon^{(2)}_{i_2\bk_2}) \prdn{2}{1} + \bn{2} \overline\bV \kn{1} \label{eq:mat3} \\
\delta \tH_{22} & = & \bn{2} \bV^{(1)} \kn{2} \label{eq:mat4},
\end{eqnarray}
and with $\overline\bV = (\bV^{(1)}+\bV^{(2)})/2$. Here $(i_1,\bk_1)$
and $(i_2,\bk_2)$ are the $\bk$-vectors and band indices of the states from the Dirac 
cones of the UBZ and RBZ that map to the BBZ Dirac cone.

Using the approach of the previous section we can now determine
when the matrix elements involved in Eqs.~(\ref{eq:mat1})-(\ref{eq:mat2}) vanish.
In particular, we know that for a 
sufficiently small misorientation $\theta$ the vanishing or
not of these matrix elements is governed
solely by the shift term $\Delta\tilde{\bG}$, i.e. by $\bk_2-\bk_1$.

To determine $\Delta\tilde{\bG}$ we must therefore specify the difference
$\bk_2-\bk_1$, that is the difference between the $\bk$ vectors
of the two single layer Dirac cone states that map back to the
same $\bk$-vector in the BBZ. If we consider the case $\delta=3$ then,
reading from Table \ref{tab:kmap3}, we find that the special $K$-point pair
$(\bK_U^{\ast},\bK_R)$, or its conjugate pair, is folded back to the same special
$K$-point in the BBZ.
Clearly $\bk$-point differences are unchanged by shifting all $\bk$-points
by some $\delta\bk$, and therefore $\bk_2-\bk_1=\bK_U^{*}-\bK_R$.

Expressing $\bK_U^{*}-\bK_R$ in coordinates of the bilayer
reciprocal lattice we then find $\bK_U^{*}-\bK_R = (q-p)/(2\gamma) (\bg_1+\bg_2)$,
i.e. that $l_1 = l_2 = (q-p)/(2\gamma)$ in Eq.~(\ref{eq:l1l2}). Using this
and substituting into the shift term of Eq.~(\ref{eq:msol3}) we find

\begin{equation} \label{eq:shft}
\Delta\tilde{\bG} = -\frac{q-p}{2p}\begin{pmatrix} 1 \\ 1 \end{pmatrix}
\end{equation}

We can easily ensure this is integer valued (as it must be)
by the choice $q=p(1+2n)$, with $n$ an integer. We thus
conclude that both
the commensuration reciprocal lattice vectors \emph{and the
shift from the origin} diverge as $q\rightarrow\infty$, that is, as
$\theta\rightarrow0^\circ$.
The first order shift will therefore be
\emph{negligible for all $q$ that result in commensuration cells
greater than some critical size}, which will depend on the
particular form of the $C_{\bk_1+\bG_1}^{\bk_2+\bR\bG_2}$, i.e. on details of the
electronic structure.

\begin{figure}
\caption{(Color online) Shown is the shift term $|\Delta\tilde{\bG}|$ corresponding to the
separations $\bk_2-\bk_1=\bK_U^{\ast}-\bK_R$ for $p = 1,2,17$ ($\delta=3$), and
$\bk_2-\bk_1=\bK_U-\bK_R$ for $p=3$ ($\delta=1$). Note that this shift diverges in 
all cases as $\theta\rightarrow0$.
\vspace{1.0cm}
}
{\includegraphics[scale=0.32,angle=00]{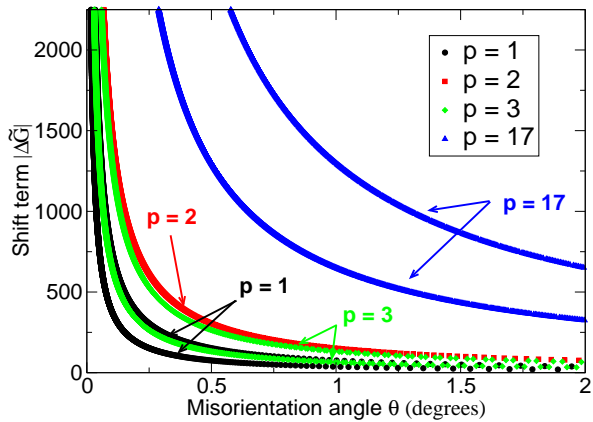}}
\label{fig:Kshft}
\end{figure}

If we reflect that as $\theta\rightarrow0$ then both $\Delta K_1=|\bK_U^{\ast}-\bK_R|\rightarrow2/3$ and
$\Delta K_2=|\bK_U-\bK_R^{\ast}|\rightarrow2/3$ while, 
at the same time, $g=|\bg_1|=|\bg_2|\rightarrow0$ (see Eq.~\ref{eq:g3}),
then it is not surprising that $\Delta\tilde{\bG}$ diverges. The behavior of $\Delta\tilde{\bG}$
in the case where $\delta=1$ is, however, not so clear. In this case $K$-points separated by
$\Delta K$ map back to the BBZ $K$-points, and so $\bk_2-\bk_1=\bK_U-\bK_R=\bK_U^{*}-\bK_R^{*}$,
and this, as may be seen from Fig.~\ref{fig:diobz} goes to zero as $\theta\rightarrow0$. In this case
neither Eq.~\ref{eq:msol1} or Eq.~\ref{eq:msol2} yield an integer valued $\Delta\tilde{\bG}$ and
so they may not be used. The Diophantine problem can, however, be solved numerically
with the result that, as shown in Fig.~\ref{fig:Kshft}, $|\Delta\tilde{\bG}|$ diverges as 
$\theta\rightarrow0$ in this case also.

The question then arises if higher order terms in perturbation theory
may lead to a \emph{splitting} of the Dirac cones. In fact,
it is easy to show that
such terms may lead only to an \emph{equal shift of both bands}.
This can be seen by an examination of the quantities
involved in higher orders of perturbation theory.
Let us first consider the calculation of the shift of the
unrotated layer Dirac band. All terms in the perturbation
expansion (which we do not need to consider explicitly)
will involve matrix elements $\bn{1} \overline\bV \kn{2}$
and $\bn{1} \bV^{(2)} \kn{1}$, and the unperturbed eigenvalues,
which are just those of single layer graphene. Now,
if we consider the shift of the rotated layer Dirac band
we see that the relevant matrix elements are either
the conjugate of those involved in the former case,
$\bn{2} \overline\bV \kn{1}=\bn{1} \overline\bV \kn{2}^{\ast}$, 
or are equal by the
symmetry of the bilayer, $\bn{2} \bV^{(1)} \kn{2}=\bn{1} \bV^{(2)} \kn{1}$.
Since the unperturbed eigenvalue spectrum is again that of
single layer graphene we immediately see that all terms in
the perturbation expansion for the eigenvalue shift of
the rotated and unrotated layers will be identical,
and hence  also the final energy shift.

Thus it is only the first order term that can break
the degeneracy of the Dirac cones from each layer and,
as we have shown above, this is zero for all
$N_{C}$ greater than some critical value.
Since it is known from \emph{ab-initio} calculations\cite{shall08} that this
degeneracy is already very small for the smallest cell $N_{C}=28$,
corresponding to $(p,q)=(1,3)$,
we can then conclude that \emph{for all commensuration cells the
Dirac bands will be effectively degenerate}, with exact
degeneracy in the $\theta\rightarrow 0^\circ$ limit or incommensurate rotations. 

When higher order terms in the perturbation expansion are unimportant
the enforced vanishing of the 1st order term therefore leads to a decoupling
of the Dirac cones. Higher orders in perturbation theory,
while unable to split the Dirac cones, may,
as we shall see subsequently, lead to a suppression of the 
the Fermi velocity of the degenerate Dirac cones and
non-linear band warping for very small misorientation angles.


\subsection{Rotation angle versus cell geometry dependence of the coupling of 
single layer states}
\label{sec:rotvpq}

\begin{figure}
\caption{(Color online) Example of the relationship between $\bk_2-\bk_1$ and
the shift of the lattice of solution vectors of the equation
$\bG_1=\bR\bG_2+\bk_2-\bk_1$. Shown in the middle panel are those
$\bk$-points that fold back to the $\Gamma$ point of the commensuration
Brillouin zone, for the case $(p,q)=(11,31)$. The left and right panels display the reciprocal
lattices $\bG_1$ and $\bR\bG_2+\bk_2-\bk_1$ along with the solution vectors
for the two different cases of $\bk_2-\bk_1$ indicated in the central panel. 
See Section \ref{sec:rotvpq} for details.
\vspace{0.5cm}
}
{\includegraphics[scale=0.32,angle=00]{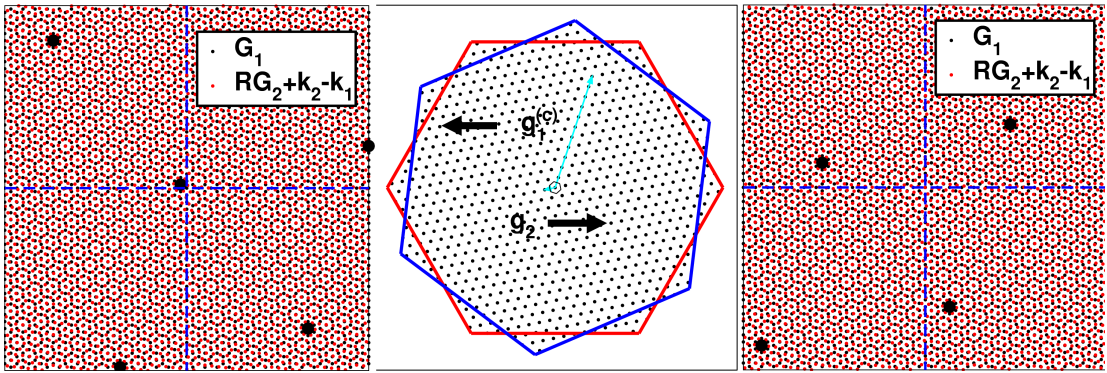}}
\label{fig:kshfts}
\end{figure}

Given a relative rotation $\theta$ between the two graphene layers there
exists an infinite set of integer pairs $(p,q)$ that, via Eq.~(\ref{eq:dioth}),
reproduce the rotation angle to arbitrary accuracy.
This includes also incommensurate
rotations, corresponding to the limit of diverging $p$ and $q$. Each of
the bilayer unit cells in this set have differing in-plane primitive
vectors $\bt_1$ and $\bt_2$ and reciprocal lattice vectors $\bg_1$ and $\bg_2$.
The \emph{symmetry allowed} coupling of two single layer states $(i_1,\bk_1)$ 
and $(i_2,\bk_2)$, governed by $\bg_1$ and $\bg_2$, can, therefore, be 
dramatically different for very similar $\theta$. As the electronic
structure of the bilayer is determined by the coupling of single layer states through
the interlayer interaction, this situation is counter-intuitive: one would expect
electronic properties to depend smoothly on the rotation angle.

Recalling the analysis of Section \ref{sec:mat}, we note that
the magnitude of the interlayer coupling is determined
by the selection rule, Eq.~(\ref{eq:comcond}), for the
coefficients of the Fourier expansion of interlayer matrix elements,
see Eq.~(\ref{eq:msum}).
As has been discussed, the coefficients 
in such a Fourier expansion decay with increasing $\bG_{1,2}$ and are
largest at or near the origin of reciprocal space. Now, the
shift term of, e.g., Eq.~(\ref{eq:msol3}) describes a \emph{scale
relation} between 
the resulting shift of allowed $\bG$-vectors in the Fourier sum,
and the $\bk$-vector difference of the coupled states
$\bk_2-\bk_1$: states coupled by $2p/\gamma(i_1\bg_1+i_2\bg_2)$ 
correspond to a shift of $-(i_1,i_2)$. Thus it is not the
$\bg_i$ that control how single layer states couple, but rather
the vectors $\bg^{(c)}_i = 2p/\gamma\bg_i$.

This is illustrated in Fig.~\ref{fig:kshfts}. Shown in the central panel
are the UBZ, RBZ, and BBZ of the  $(p,q)=(11,31)$ twist bilayer
(rotation angle $\theta= 23.16^\circ$). Indicated are two different
possible couplings of single layer states: (i) $\bk_2-\bk_1=\bg_2$, and
(ii) $\bk_2-\bk_1=11\bg_1=\bg^{(c)}_1$. 
The solution vectors of $\bG_1=\bR\bG_2+\bk_2-\bk_1$
corresponding to each of these cases are shown in the left and right hand
side panels respectively. The single layer states coupled by $\bg^{(c)}_1$
have contributing vectors close to the origin indicating a strong coupling, 
while the states coupled by $\bg_2$ have contributing vectors far from
the origin, indicating a comparatively weaker coupling. 

The extent to which states connected by $\bg^{(c)}_i$ dominate other symmetry
allowed couplings depends on the nature of the decay of the
Fourier coefficients $C_{\bk_1+\bG_1}^{\bk_2+\bR\bG_2}$, i.e. on 
details of the electronic structure. For graphene, as we show in Section IV,
the situation is of a coupling dominated entirely by the vectors
$\bk_2-\bk_1= n_1 \bg^{(c)}_1 + n_2 \bg^{(c)}_2$ with $0 \le n_i \le 1$.
Regardless of the particular details of the Fourier coefficients,
allowing $|\bt_i|\rightarrow\infty$ and $|\bg_i|\rightarrow0$ for
a given rotation angle $\theta$, beyond a certain point, introduces no new interlayer
coupling to the bilayer system as all additional symmetry allowed couplings will
have zero magnitude by the arguments above. Thus the ultimate smoothness
of electronic properties with $\theta$ is guaranteed by the fact of
the decay the Fourier coefficients
$C_{\bk_1+\bG_1}^{\bk_2+\bR\bG_2}$, a natural result.
Clearly, the faster the decay of these coefficients
the greater the dominance of a few $\bg^{(c)}_i$ in the coupling of single layer states and,
hence, the less the electronic structure depends on details of the real space cell.

To determine a general form of the $\bg^{(c)}_i$,
we first calculate $g=|\bg_i|$ which, for $\delta=1$ is given
by

\begin{equation} \label{eq:g1}
g = \frac{2\gamma}{\sqrt{9(p^2+3q^2)}} =
 \frac{2\gamma}{3p}\sin\theta/2 = \frac{2\gamma}{3\sqrt{3}q}\cos\theta/2.
\end{equation}
and for $\delta = 3$ by

\begin{equation} \label{eq:g3}
g = \frac{2\gamma}{\sqrt{3(p^2+3q^2)}} =
 \frac{2\gamma}{\sqrt{3}p}\sin\theta/2 = \frac{2\gamma}{3q}\cos\theta/2.
\end{equation}
To determine $g^{(c)}=|\bg^{(c)}_i|$ we then multiply $g$ by the
inverses of the pre-factors to the shift terms in Eqs.~(\ref{eq:msol1})-(\ref{eq:msol4}),
i.e. by $6q/\gamma$, $6\sqrt{3}p/\gamma$, $2p/\gamma$, and $6\sqrt{3}q/\gamma$
respectively
(the factors of $\sqrt{3}$ arise from the different $(l_1-2l_2,2l_1-l_2)^{T}$ structure of 
Eqs.~(\ref{eq:msol2}) and (\ref{eq:msol4})). In this way we find that the relevant $g^{(c)}$ is
given by

\begin{equation} \label{eq:gc}
g^{(c)} = \frac{4\sqrt{3}}{\delta}\sin\frac{\theta}{2},
\end{equation}
and that $\bg^{(c)}_i = g^{(c)}\hat{\bg_i}$ with $\hat{\bg_i}$ the unit
vectors formed from the primitive vectors $\bg_i$. As expected,
these coupling vectors depend only on the misorientation angle.
Interestingly, $g^{(c)}$ may be expressed in terms of $\Delta K$,
i.e. in terms of the separation of pairs of special K-points $(\bK_U,\bK_R)$,
see Fig.~\ref{fig:diobz}, as

\begin{equation} \label{eq:coupK}
g^{(c)} = \frac{3\sqrt{3}}{\delta} \Delta K.
\end{equation}
In Ref.~[\onlinecite{san07}] it was noted that $\Delta K$ constitutes an energy
scale of the twist layer physics. Here, in our more general lattice treatment,
we see that this is also an important reciprocal space length scale, describing
the coupling of single layer states by the interaction.


\subsection{Reduction in Fermi velocity of the Dirac cones}

Here we examine the impact upon the degenerate Dirac cones of the
bilayer of the non-zero matrix elements. Let us consider
the Dirac cone state from the unrotated layer situated at $\bK_U+\delta\bk$ 
which, therefore, has eigenvalue

\begin{equation}
\epsilon^{(1)} = s_1|\delta\bk|
\end{equation}
where we set $\hbar v_F = 1$, and $s_1$ is a sign indicating whether
the eigenvalue belongs to the electron or hole Dirac cone. From 
Section \ref{sec:mat} we know
that this state will couple with the two states from the rotated layer Dirac cone
also having $\bk$-vector $\bK_U+\delta\bk$. These states have energies

\begin{equation}
\epsilon^{(2)} = s_2|\delta\bk-\Delta{\bf K}|
\end{equation}
In this equation $\Delta\bK$ is a reciprocal space vector connecting
nearest neighbor Dirac cones (for $\theta < 30^\circ$ that we consider here), 
see Fig.~\ref{fig:diobz}.
Due to the translational symmetry, there may also be coupling
with states at $\bK_U+\delta\bk-n_1\bg_1^{(c)}-n_2\bg_2^{(c)}$ which have
energies

\begin{equation}
\epsilon^{(2)}_{n_1 n_2} = s_2|\delta\bk-\Delta{\bf K} 
 - n_1\bg_1^{(c)} - n_2\bg_2^{(c)}|
\end{equation}
where $n_{1,2}$ are integers.
Note the use the coupling vectors $\bg_i^{(c)}$,
that determine which single layer states couple through the 
interlayer interaction, see Section \ref{sec:rotvpq}.
Choosing $\delta=3$ we find from Eq.~\ref{eq:coupK}
that

\begin{eqnarray}
\bg_1^{(c)} & = & \sqrt{3}\Delta K \hat{\bf g}_1 \label{eq:ggc1} \\
\bg_2^{(c)} & = & \sqrt{3}\Delta K \hat{\bf g}_2 \label{eq:ggc2}
\end{eqnarray}
with $\hat{\bf g}_1$ and $\hat{\bf g}_2$ the unit vectors
of the bilayer reciprocal lattice. 
Using this one may then show that

\begin{equation} \label{eq:Kbit}
|\Delta{\bf K} + n_1\bg_1^{(c)} + n_2\bg_2^{(c)}| = \Delta K \eta_{n_1 n_2}
\end{equation}
where

\begin{equation} \label{eq:eta}
\eta^2_{n_1 n_2} = 1 + n_1 n_2 + 3(n_1-n_2)(n_1-n_2-1)
\end{equation}
i.e. that the $\theta$ dependence of 
$|\Delta{\bf K} + n_1\bg_1^{(c)} + n_2\bg_2^{(c)}|$ is \emph{entirely
through $\Delta K$}. We may then expand 
$\epsilon^{(2)}_{n_1 n_2}$ as

\begin{equation} \label{eq:ep2}
\left(\epsilon^{(2)}_{n_1 n_2}\right)^2 = \delta k^2 + \Delta K^2\eta_{n_1 n_2}^2
-2\delta k\Delta K\eta_{n_1 n_2}\cos\phi_{n_1 n_2}.
\end{equation}
In this expression 
$\phi = \angle (\delta\bk,\Delta{\bf K} + n_2\bg_1^{(c)} + n_2\bg_2^{(c)})$,
and $\delta k = |\delta\bk|$. The overall eigenvalue shift may be written
to second order as

\begin{equation} \label{eq:ep}
\delta\epsilon = \sum_{n_1 n_2} \left\{
\frac{\alpha_{n_1 n_2}^{s_1-}}{\epsilon^{(1)}+\epsilon^{(2)}_{n_1 n_2}}
+ \frac{\alpha_{n_1 n_2}^{s_1+}}{\epsilon^{(1)}-\epsilon^{(2)}_{n_1 n_2}}
\right\}
\end{equation}
where $\alpha_{n_1 n_2}^{s_1-}$ and
$\alpha_{n_1 n_2}^{s_1+}$ are coupling constants between the different
states (i.e., squares of overlap elements
of the kind discussed in Sections III C and III D).
Finally, from Eqs.~(\ref{eq:Kbit}-\ref{eq:ep}) 
may then be derived that this shift leads to a reduction in
the Fermi velocity given by

\begin{equation} \label{eq:fv}
v_F = v_F^{(SL)}(1 - \frac{\alpha}{\Delta K^2})
\end{equation}
where $v_F$ is the Fermi velocity of the misoriented layers,
$v_F^{(SL)}$ the Fermi velocity of SLG, and $\alpha$ an overall coupling
constant given by

\begin{equation}
\alpha = \sum_{n_1 n_2} \frac{\alpha_{n_1 n_2}^{s_1-} + \alpha_{n_1 n_2}^{s_1+}}
 {\eta_{n_1 n_2}^2}.
\end{equation}
Thus $\alpha>0$ as required for Eq.~\ref{eq:fv} to actually describe
a reduction in the Fermi velocity.

The angle dependence contained in $\Delta K$ can, equivalently,
be expressed via the moir\'e periodicity (see Section II) $D$ leading to
the form

\begin{equation} \label{eq:fv1}
v_F = v_F^{(SL)}(1 - \beta D^2)
\end{equation}
where $\beta$ is a related coupling constant.

Note that while Eq.~(\ref{eq:fv}) is of the same form as that derived by
Santos \emph{et al.} in Ref.~[\onlinecite{san07}], the origin is somewhat different.
Rather than use a continuum approximation, as deployed in Ref.~[\onlinecite{san07}],
we have retained the lattice physics which, in fact, only
enters in the form of the 'coupling' primitive vectors, Eqs.~(\ref{eq:ggc1})
and (\ref{eq:ggc2}).


\section{Tight-binding analysis}

\begin{figure}
\caption{(Color online) Tight-binding calculation
of the Dirac point splitting in (i) AB bilayer, indicated by light shaded points and
(ii) $\theta=30^\circ\pm8.21$ twist bilayers, indicated by dark shaded open squares
($\theta=38.21^\circ$) and diamond symbols ($\theta=21.79^\circ$). The \emph{ab-initio}
values for the AB and twist bilayer splitting are given by the horizontal dot-dashed
lines.
Full/dashed lines are calculations with the environment dependence of the hopping
matrix elements switched off/on.
\vspace{1.5cm}
}
{\includegraphics[scale=0.50,angle=00]{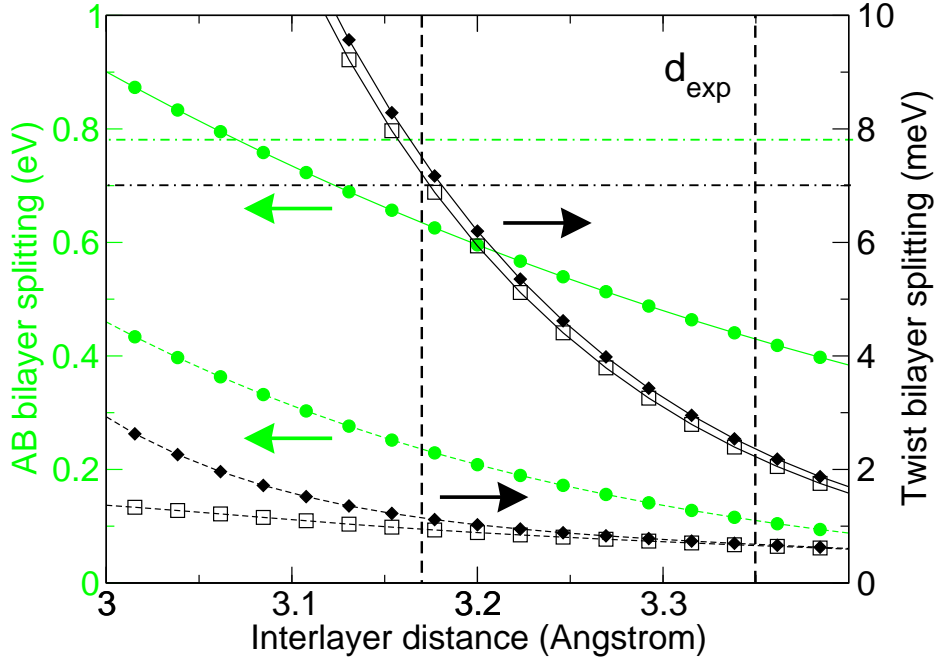}}
\label{fig:dps}
\end{figure}

We now turn to tight-binding calculations of the twist bilayer
structures elucidated in Section II. Given that the number
of atoms in the real space commensuration cell diverges
as the rotation angle $\theta\rightarrow0^\circ$ the tight-binding
method offers perhaps the only way of exploring this interesting
limit; \emph{ab-initio} calculations are certainly not practical.
Here we shall employ the same tight-binding method deployed by
Santos \emph{et al.} in their continuum approach to
the twist bilayer\cite{san07}, one of the so-called environment dependent
tight-binding methods\cite{tang96}.

In the original article by Tang \emph{et. al} the environment
dependent parameterization for carbon that these authors
proposed was checked against a database including, amongst other 
three dimensional lattices, diamond and graphite, as well 
as a one dimensional
carbon chain. Unfortunately, graphene and graphene-based
structures were not part of this dataset, and it is thus
important to first verify the accuracy of this method for
this case. A sensitive test of accuracy is provided by the
Dirac point splitting of graphene bilayer structures,
which may be quite large in the case of the AB stacked
bilayer (the \emph{ab-initio} value is 0.78~eV) and
on the other hand rather small in the case of twist bilayers e.g.
7~meV for both the $\theta=30^\circ\pm8.21^\circ$ twist 
bilayers\cite{shall08}. This latter case entails
a particularly sensitive test as, although both the
$\theta = 38.21^\circ$ and $\theta = 21.79^\circ$ systems
have the same Dirac point splitting, the crystal geometries
are actually quite different\cite{shall08}.

In Fig.~\ref{fig:dps} are shown calculations of the Dirac point splitting
for the AB bilayer, as well as the two twist bilayers with 
$\theta=30^\circ\pm8.21^\circ$. Surprisingly, one finds
that even the Dirac point splitting of the AB bilayer is not well
reproduced; a much reduced interlayer separation is required
to recover the \emph{ab-initio} result of 0.78~eV. In addition,
the splitting of the $\theta=30^\circ\pm8.21^\circ$ twist
bilayers is also underestimated and, furthermore, is quite
different between the $\theta=38.21^\circ$ and 
$\theta=21.79^\circ$ cases.

Fortunately, this situation is significantly improved by switching
off the environment dependence of the hopping integrals,
in which case the method is simply the usual tight-binding
scheme with distance dependent pairwise hopping matrix elements
(see Ref.~[\onlinecite{tang96}]). Given this, a
reasonable agreement with \emph{ab-initio} calculations
may be found. For a somewhat reduced interlayer distance of
3.17~\AA (5\% smaller than the nominal experimental interlayer distance
of 3.34~\AA), we find $\approx7$~meV for the twist bilayer
splitting, which is in very good agreement with \emph{ab-initio} data,
and a splitting of $0.63$~eV for the AB bilayer, which is less 
good but still reasonable. This interlayer distance is indicated
by a dashed vertical line in Fig.~\ref{fig:dps}.
Our choice of calculation method is therefore the
parameterization of Tang \emph{et al.}, but with the environment dependence
suppressed, and the interlayer distance set to 3.17~\AA.

\begin{figure}
\caption{(Color online) Tight-binding band structures for rotation angles of
$\theta=21.79^\circ$ (1,3), $9.43^\circ$ (1,7), $2.65^\circ$ (1,25), 
$1.47^\circ$ (1,45), displayed respectively clockwise from top left.
Shown is the band structure generated by
direct tight binding calculation (wide black lines) and that generated with SLG basis 
approach, indicated by light shaded (green) lines. For comparison in each panel 
the folded back band structure of single layer graphene is shown (dashed lines).
The numbering of the panels corresponds to that of Fig.~\ref{fig:diobz}.
\vspace{1.5cm}
}
{\includegraphics[scale=0.50,angle=00]{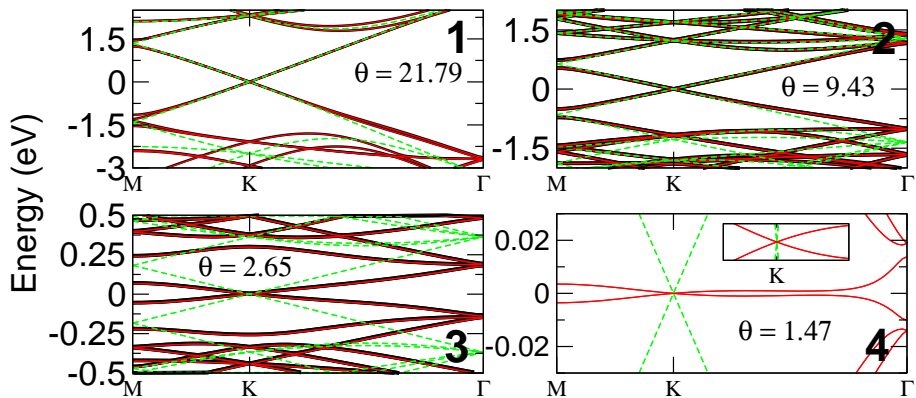}}
\label{fig:band}
\end{figure}

While the tight-binding method extends the domain of direct band structure
calculation far beyond that which may be achieved by the \emph{ab-initio}
approach, for the particular case of bilayer graphene (or more
generally graphene stacks) one may do significantly better. In
particular, if it is only the low energy 
band structure that is of interest, the single layer graphene
basis introduced in Section III is much more appropriate
that the full tight-binding basis set. In this basis
the bilayer Hamiltonian is

\begin{equation} \label{eq:m2}
[\bH(\bk)]_{i_1\bk_1 i_2\bk_2} = \begin{pmatrix}
 \bi{i_1\bk_1}{1} \bH \ki{i_2\bk_2}{1} & \bi{i_1\bk_1}{1} \bH \ki{i_2\bk_2}{2} \\
 \bi{i_1\bk_1}{2} \bH \ki{i_2\bk_2}{1} & \bi{i_1\bk_1}{2} \bH \ki{i_2\bk_2}{2} \\ 
\end{pmatrix},
\end{equation}
where $i_1\bk_1$ and $i_2\bk_2$ are the band- and $\bk$-indices of 
states that fold back to $\bk$ (a reciprocal lattice vector in the BBZ),
and the superscript of the kets has the same meaning as in Section III,
i.e., $(1)$/$(2)$ refers to eigenkets of the unrotated/rotated layers.
Eigenvalues at $\bk$ may then be obtained by diagonalising the
matrix consisting of all states $i_1\bk_1$ and $i_2\bk_2$ that fold
back to $\bk$.

Matrix elements in Eq.~(\ref{eq:m2}) will involve both on-site 
terms and terms involving interlayer
hopping integrals $\bn{n'} \overline{\bV} \kn{n}$ where
$\overline{V} = \frac{1}{2}(\bV^{(1)}+\bV^{(2)})$.
The matrix elements $\bn{1} \bV^{(2)} \kn{1}$ and $\bn{2} \bV^{(1)} \kn{2}$ are 
equivalent to three center hopping integrals, and so may be set to zero,
while the matrix element $\bn{1} \overline{\bV} \kn{2}$ may be evaluated as

\begin{equation} \label{eq:tbmat}
\gb{\phi^{(1)}_{i_1\bk_1}} \overline{\bV} \gk{\phi^{(2)}_{i_2\bk_2}} = \frac{1}{N_{C}}
\sum_{n_1 n_2} e^{i\bk_2.\bR_{n_2}} e^{-i\bk_1.\bR_{n_1}} 
 a^{p_z}_{i_1\bk_1} a^{p_z}_{i_2\bk_2} \left(n_z^2t_{pp\sigma} + (1-n_z^2)t_{pp\pi}\right).
\end{equation}
where $\bR_{n_1}$ and $\bR_{n_2}$ are vectors from layer 1 
(unrotated) and layer 2 (rotated) respectively,
the sum is over all atoms in the twist boundary primitive cell, $a^{p_z}_{i_n\bk_n}$
is the $p_z$-coefficient of the eigenvector corresponding 
to the $i_n\bk_n$ state from layer n, $n_z$ a directional cosine, $t_{pp\sigma}$ and
$t_{pp\pi}$ distant dependent hopping integrals, and $N_C$ the number of carbon atoms
in the bilayer primitive cell.

\begin{figure}
\caption{(Color online) Convergence of two nearly degenerate eigenvalues on the lower branch of the bilayer 
Dirac cone. The rotation angle of the bilayer is $\theta=3.48^\circ$ 
(corresponding to $(p,q)=(1,19)$ - see section II). Eigenvalues are calculated at
$\bk = \bK + \frac{2}{10}({\bf\Gamma}-\bK)$ with $\bK$ and ${\bf \Gamma}$ the $\bk$-vectors
of the K and $\Gamma$ points. Inset shows $\log_{10}|\bn{1} \overline{\bV} \kn{2}|$
plotted for $\bk_2-\bk_{\Gamma} = n\bg_1$.
\vspace{1.5cm}
}
{\includegraphics[scale=0.50,angle=00]{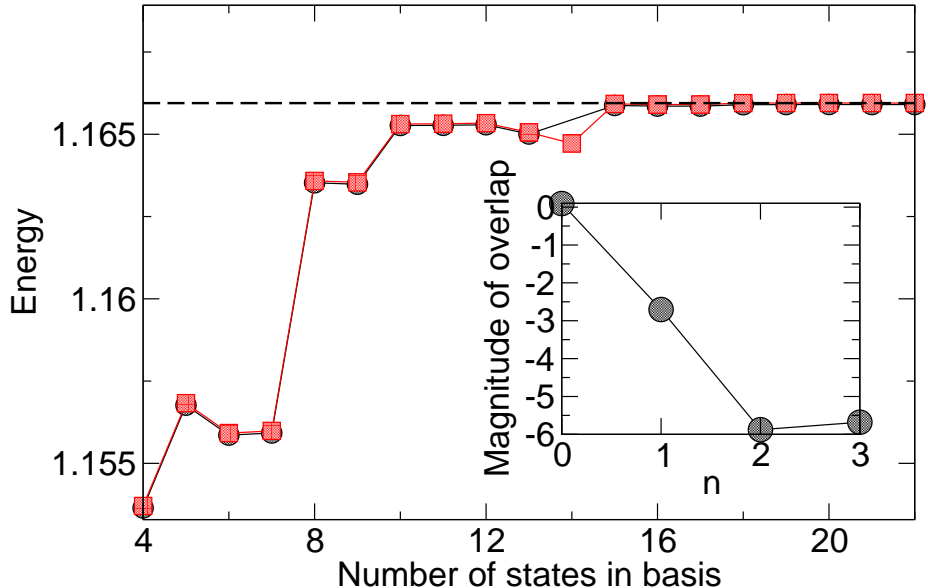}}
\label{fig:conv}
\end{figure}

The advantage of this approach is that if it is
only the low energy band structure that is of interest, then
only low energy eigenkets of the unrotated and rotated layers
are needed in constructing the SLG basis. In practice,
the number of such low energy states required
will depend on the strength of
the interlayer interaction, which for graphene layers
is weak and leads to a rather rapid convergence of
the basis set, as shown in Fig.~\ref{fig:conv}. For the
case of the $\theta=3.48^\circ$ bilayer shown the maximum
size of the basis set is 2048 states, and numerical convergence
is reached at 24 states.
A further advantage for the special case of twist
bilayers lies in fact that from the selection rule
analyzed in Section \ref{sec:mat}
one knows that \emph{a priori}
many of the matrix elements in Eq.~(\ref{eq:m2}) will
be zero. Inspection of the numerical value of
the interlayer matrix elements, Eq.~(\ref{eq:tbmat}),
shows that they are negligible for
$\bk_2-\bk_1= n_1 \bg^{(c)}_1 + n_2 \bg^{(c)}_2$ 
with $n_i > 1$ (see Section \ref{sec:rotvpq}
for a description of $\bg^{(c)}_{1,2}$); a typical
case is illustrated in the inset of Fig.~\ref{fig:conv}.

For actual calculations one may then utilize a truncated basis
in which only a fraction of the actual matrix elements
required need be calculated. This extends by more than an
order of magnitude the number of carbon atoms that may
be considered: within the SLG approach $N_C=26,068$ could
be treated within the same time that, by direct tight-binding calculations,
a system of $N_C\approx 1200$ could be calculated.

\begin{figure}
\caption{(Color online) Relation between misorientation angle of the bilayer and Fermi
velocity damping; dashed (black) line in the main panel is the best fit
to Eq.~(\ref{eq:fv}) for data points with $\theta>5^\circ$.
\vspace{1.5cm}
}
{\includegraphics[scale=0.50,angle=00]{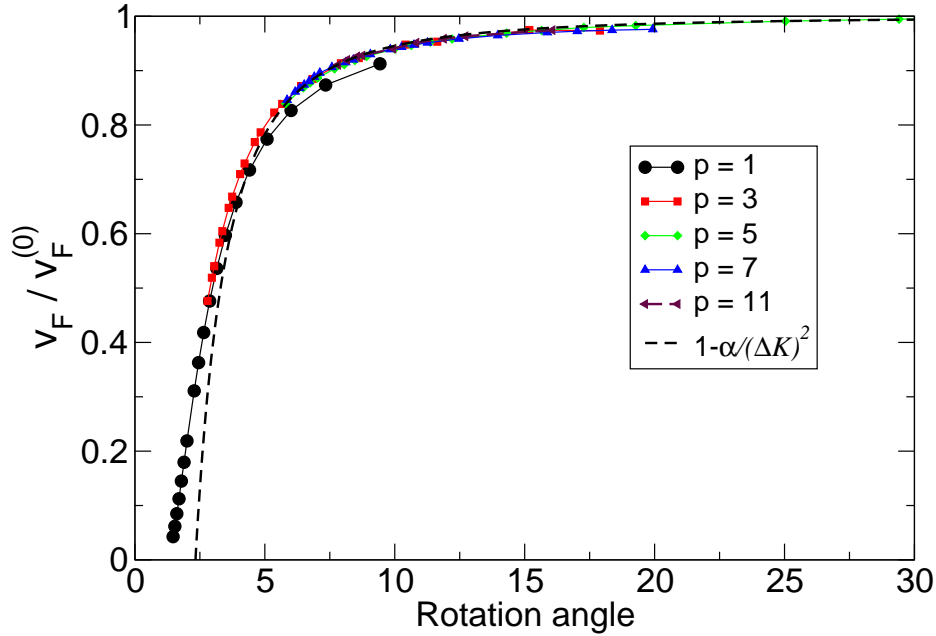}}
\label{fig:fv}
\end{figure}

\begin{figure}
\caption{(Color online) Splitting of bands at the Dirac point in misoriented 
graphene layers. Shown are tight-binding binding calculations for supercells generated with
$p=1$ and $q$ an odd integer. First principles calculations for supercells generated
by $(p,q)$ pairs of (1,3), (1,2), (1,5), (2,3), (1,7) in order of increasing $N$
are taken from Ref.~[\onlinecite{shall08}].
\vspace{1.5cm}
}
{\includegraphics[scale=0.50,angle=00]{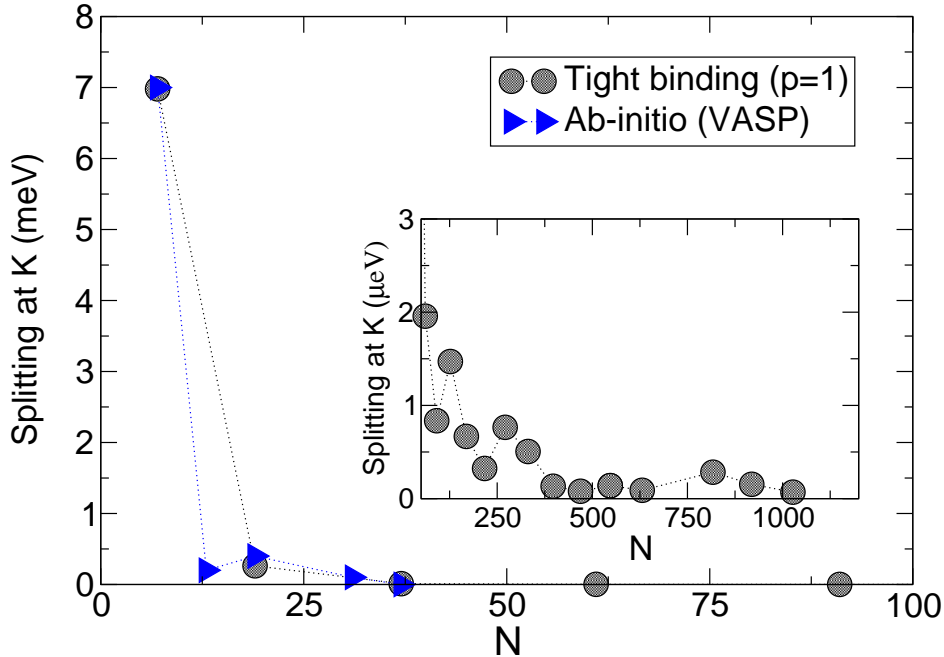}}
\label{fig:ds}
\end{figure}

We now consider the low energy electronic structure of a selection of
twist bilayers, shown in Fig.~\ref{fig:band}, with the band structure
plotted along the $\text{MK}\Gamma$ high symmetry points path in the BBZ. 
In panels 1-4 are shown 4 twist bilayers in the set
$p=1$, $q\in\text{odd}\,\mathbb{Z}$ with $q = 3,7,25,45$ (misorientation 
angles of $\theta = 21.79^\circ$, $9.43^\circ$, $2.65^\circ$, 
and $1.47^\circ$ respectively). In panels 1-3 are shown band structures
generated by both direct tight-binding calculation as well as
the SLG basis outlined above; clearly these two approaches lead to
identical results, as expected. A number of interesting features may be 
noted from these band structures. Firstly, as expected from the
general analysis of Section III the Dirac point always decouples, i.e.,
there is no splitting of the degenerate Dirac bands from each layer,
see Fig.~\ref{fig:ds}.
On the other hand, in agreement with \emph{ab-initio} calculations
\cite{hass08,lat07,shall08},
one observes that \emph{away} from the Dirac point the bilayer band structure
clearly shows a perturbation due to layer interaction, as may be
observed in Fig.~\ref{fig:band} by the strong band hybridization at
the $\Gamma$ point.
This, as discussed in Section \ref{sec:mat}, is due to the term
$\bk_2-\bk_1$ in the selection rule
$\bG_1 = \bR \bG_2 + \bk_2-\bk_1$; when $\bk_2-\bk_1=0$ 
the interlayer matrix elements never vanish leading to a strong
coupling for these states.
In addition, one notes a reduction in Fermi velocity
of the Dirac cone as the rotation angle $\theta\rightarrow0^\circ$.

In Fig.~\ref{fig:fv} is shown this Fermi velocity suppression as a function
of the rotation angle of the twist bilayer. Clearly, this effect is quite 
substantial, and for $\theta=1.47^\circ$ (the smallest angle 
calculated) $v_F$ is only 5\% of the value of SLG. Interestingly,
for $\theta>5^\circ$ the tight-binding data is very well described
by Eq.~(\ref{eq:fv}), in the small angle limit, however, the
failure of any fitting of the form Eq.~(\ref{eq:fv}) indicates the
importance of higher orders of perturbation theory. Clearly, the
band structure in the $\theta\rightarrow0^\circ$ limit is profoundly
altered from that of SLG with, however,
the degeneracy of the Dirac bands from each layer preserved,
as was proved must be the case in Section III D. (One should note
that, in contrast to the first principles calculation of
Ref.~\cite{shall08}, we find a small reduction in Fermi velocity
for the case of $\theta=9.43^\circ$; such a difference presumably 
reflects the fact that parameterisation of Tang et al.\onlinecite{tang96},
not explicitly designed for low energy graphene physics,
has scope for improvement.)

As has been mentioned, this preservation of the degeneracy is 
a striking illustration of the singular nature of the $\theta\rightarrow0^\circ$
limit; for any finite $\theta$ one has a \emph{fourfold} degeneracy
at the Dirac point,
while at $\theta=0$ one has the AB bilayer with a twofold
degeneracy at the Dirac point, and with the other Dirac bands bands split
by $0.78$ eV.
An interesting question, which we shall only pose here and not answer,
is how this Fermi velocity reduction would be altered by both charge
self-consistency and many-body effects. Both of these may be expected to
become more important as the Fermi velocity is reduced, and may
dramatically change the nature of the small angle electronic structure.

Graphene stacks grown on the C-face of SiC typically have Fermi velocity
reductions of 20-30\% which then implies, assuming that such a reduction is
entirely due to rotation, misorientations of $\theta>5^\circ$. Given that the
formation energy of a twist bilayer
increases as $\theta\rightarrow0^\circ$\cite{shall08a}, 
with the minimum defect energy for $30^\circ\pm2.20^\circ$\cite{shall08a},
it makes sense that, on average, misorientation angles with $\theta<5^\circ$ should play
a less important role.
On the other hand, it is clear that samples with $\theta<5^\circ$
do exist; STM experiments detect moir\'e patterns that correspond to angles in the
$1.9^\circ-19^\circ$ range\cite{varch08}. 
In contrast, for a graphene slab dominated by 
$30^\circ\pm2.02^\circ$ rotations, or simply large angle rotations (such a system was studied 
in Ref.~[\onlinecite{hass08}], see also Ref.~[\onlinecite{sprin09}])
one would expect to have, in addition to a Dirac spectrum over a wide
energy range, a Fermi velocity exactly that of SLG.


\section{Conclusions}

To conclude we have given a complete description of the
possible commensurations of graphene layers misoriented by some angle
$\theta$. We find that the condition for a commensuration to occur is that,
expressed in lattice coordinates of the unrotated layer, the rotation matrix 
connecting the layers be rational valued, and thus the complete set of commensurations
is described by two integers, which we denote $(p,q)$. For any such
bilayer, we have shown that the $K$ points of the unrotated and rotated layers
map directly to $K$ points of the bilayer Brillouin zone; a fact that plays
an important role in the interlayer interaction.

We have further shown that the nature of the interlayer
interaction may be understood by a $\bk$-dependent interference
condition, that may be expressed as a commensuration of the
single layer reciprocal lattices.
This guarantees the decoupling of the
Dirac point and the degeneracy in the Dirac cones from
each layer, but does not preclude interactions between all
states. These latter interactions in fact give rise to a reduction
of the Fermi velocity in the $\theta\rightarrow0^\circ$ limit, and
we find a form of this Fermi damping which agrees with that
presented by Santos \emph{et al.}\cite{san07}, although
our derivation is independent of any continuum approximation.
As an interesting consequence of this analysis,
we are able to show that the bilayer electronic structure
will, in general, depend only on the misorientation angle of the
layers and not on the details of the real space unit cell.

To complement this general analysis we have calculated band structures of a wide range 
of graphene twist bilayers via the tight-binding method. By the introduction of a basis 
of single layer graphene eigenkets, which we show to be significantly more
efficient for the case of the twist bilayer, we are able to probe the
band structure in the small angle limit with relative
computational ease. We find Fermi velocity reduction that, 
for rotation angles in the range $5^\circ<\theta<30^\circ$ agrees very 
well with the form presented here and in Ref.~[\onlinecite{san07}], but that
for $\theta<5^\circ$, where at $\theta=1.47^\circ$ the Fermi velocity is only 5\% of the 
SLG value, the reduction cannot be described in this way. 
In fact, the Dirac bands in the small angle limit, while guaranteed to be degenerate,
show non-linear distortion away from the Dirac point. Thus the graphene
twist bilayer encompasses a wide range of electronic behavior, from
essentially SLG behavior for large angle rotations to quite different behavior
in the small angle limit which, nevertheless, shares important features with the
large angle case. While small angle rotations\cite{varch08} (as low as $1.9^\circ$)
have been observed experimentally, the electronic
properties of such low angle misoriented layers has yet to be experimentally
explored. The possibility of a graphene type behavior different from both SLG and the AB bilayer
makes interesting the further study of this low angle limit.

\vspace{0.5cm}

The authors acknowledge Deutsche Forschungsgemeinschaft for financial support,
and gratefully recognize collaboration within the Interdisciplinary Centre for
Molecular Materials at the University of Erlangen-N\"urnberg.

\appendix

\section{Derivation of the commensuration lattice primitive vectors}

In this Appendix we wish to determine the primitive lattice vectors
of the commensuration lattice given by Eq.~(\ref{eq:dio6a}), i.e., the
lattice defined by

\begin{equation}
\bem = \alpha \begin{pmatrix}  -p+3q \\ 2p \end{pmatrix}
 + \beta \begin{pmatrix}  -2p \\  p+3q \end{pmatrix}
\end{equation}

We first notice that only coprime $p,q$ correspond
to unique solutions. Given this, a necessary condition for
recovering lattice vectors is obviously the elimination of
the greatest common divisor (gcd) of the components of all the
(integer valued) vectors in Eqs.~(\ref{eq:dio6a}) and (\ref{eq:dio6b}).
This leads to following commensuration primitive vectors:

\begin{equation} \label{eq:pv1a}
\bt_1 = \frac{1}{\gamma} \begin{pmatrix} p + 3q \\ - 2p \end{pmatrix},
\bt_2 = \frac{1}{\gamma} \begin{pmatrix} 2p \\ -p+3q \end{pmatrix}
\end{equation}
with $\gamma = \GCD(3q+p,3q-p)$, and where we have used the fact that
$\GCD(x,y)=\GCD(x+cy,y)$ with $c$ an arbitrary integer. Possible values of $\gamma$ involve
the additional parameter $\delta = 3/\GCD(p,3)$, and are displayed in
Table~\ref{tab:gam}.

To prove that these are indeed the primitive vectors requires
further that there is no linear combination of them yields
\emph{integer valued vectors} of smaller length. In fact, as we will
now show, the vectors given in Eq.~(\ref{eq:pv1a}) are primitive only for the
case where $\delta=1$, and that when $\delta=3$ it is the linear
combinations $1/3(-\bt_1+2\bt_2)$ and $1/3(-2\bt_1+\bt_2)$ that
form the primitive vectors of the commensuration lattice.

We first take a linear combination of the supposed primitive
vectors as follows:

\begin{equation} \label{eq:abdef}
\frac{\alpha_i}{N_i} \bt_1 + \frac{\beta_i}{N_i} \bt_2 = \bt_i'
\end{equation}
where $\alpha_i,\beta_i,N_i\in\mathbb{Z}$ and the index $i=1,2$.
By eliminating common factors of $\alpha_i$ and $\beta_i$ from $N_i$
we may choose $\GCD(\alpha_i,\beta_i)=1$.
Defining $t=|\bt_1|=|\bt_2|$ and $t'=|\bt_1'|=|\bt_2'|$,
$\bt_1$ and $\bt_2$ are primitive only if there
exists no $\alpha_i,\beta_i,N_i$ such that $t'<t$. Suppose
that for a given set of these parameters $t'>t$ then we
may write

\begin{equation} \label{eq:LC}
N_i t < N_i t' = |\alpha_i \bt_1 + \beta_i \bt_2| \le (|\alpha_i| + |\beta_i|)t
\end{equation}
so that if $t'>t$ then $|\alpha_i| + |\beta_i|>N_i$ and hence
$|\alpha_i| + |\beta_i| < N_i$ implies $t'<t$. Thus if there
exist $\alpha_i,\beta_i,N_i$ such that $|\alpha_i| + |\beta_i| < N_i$
\emph{and} the linear combinations Eq.~(\ref{eq:abdef}) remain integer
valued then the supposed primitive
vectors $\bt_1$ and $\bt_2$ are in fact not primitive.

From Eq.~(\ref{eq:pv1a}) and Eq.~(\ref{eq:abdef}) we find

\begin{eqnarray}
\left[\frac{(\alpha_i + 2\beta_i)}{\gamma N_i} p 
   + \frac{3\alpha_i}{\gamma N_i} q\right] & = & z_1 \label{eq:n1} \\
-\left[\frac{(2\alpha_i + \beta_i)}{\gamma N_i} p 
   - \frac{3\beta_i}{\gamma N_i} q\right] & = & z_2 \label{eq:n2}
\end{eqnarray}
where $z_1,z_2\in\mathbb{Z}$ in order that $\bt^{'}_i$ be integer
valued. As $\GCD(\alpha_i,\beta_i)=1$
and $\GCD(p,q)=1$, then for $z_{1,2}$ to be integer valued as claimed
we require $\gamma N_i$ to be a common factor of the coefficients of $p$ and $q$ on 
the left hand sides of Eqs.~(\ref{eq:n1}) and (\ref{eq:n2}).
Using the fact that $\GCD(\alpha_i,\beta_i)=1$, we find that possible
values of $\GCD(3\alpha_i,\alpha_i+2\beta_i)$ and 
$\GCD(3\beta_i,2\alpha_i+\beta_i)$ are 1,2,3,6.
Therefore, as the minimum possible value of $\gamma=1$,
then the maximum possible value of $N_i$ is 6. 
As $|\alpha_i| + |\beta_i| < N_i$ this in turn restricts the
possible values of $\alpha_i$, $\beta_i$. 


We have used the requirement that $\bt^{'}_i$ be integer valued to
place a condition $\gamma N_i \le 6$, not $N_i \le 6$, and hence we need to 
eliminate $(\alpha_i,\beta_i,N_i)$ that lead to non-integer $\bt^{'}_i$.
From the full set of $\{(\alpha_i,\beta_i,N_i)\}$ this then
leaves only the cases $(\alpha_i,\beta_i,N_i) = \{(1,1,3),(-1,2,3),(-2,1,3)\}$,
and using these latter two we finally find the new primitive vectors

\begin{equation} \label{eq:pv3a}
\bt_1 = \frac{1}{\gamma} \begin{pmatrix} -p-q \\ 2q \end{pmatrix},
\bt_2 = \frac{1}{\gamma} \begin{pmatrix} 2q \\ -q+q \end{pmatrix},
\end{equation}

However, for the case $\delta=1$ then we have $3\mid\gamma$ and $3\mid p$ and since $\GCD(p,q)=1$ then
$3\nmid q$ and $\bt_1$ and $\bt_2$ given by Eq.~(\ref{eq:pv3a}) cannot be integer valued. Hence for 
$\delta=1$ then Eq.~(\ref{eq:pv1a}) are already the primitive
vectors of the commensuration lattice. On the other hand if $\delta=3$ then
if both $p$ and $q$ are odd then $\gamma=2$ and $\bt_1$ and $\bt_2$ given by 
Eq.~(\ref{eq:pv3a}) are
integer valued while if one of $p,q$ is even then $\gamma=1$ and again
this is so. 

To summarize we find that for the case $\delta=1$ the the vectors
given by Eq.~(\ref{eq:pv1a}) are already primitive, while for the case $\delta=3$ 
instead Eq.~(\ref{eq:pv3a}) gives the primitive vectors.

\section{Solution of equation $\bG_1 = \bR \bG_2 + (\bk_2-\bk_1)$}

We wish to determine the solutions to the equation

\begin{equation} \label{eq:Gcom}
\bG_1 = \bR \bG_2 + (\bk_2-\bk_1).
\end{equation}
This represents a similar equation to the real space
commensuration equation, but with
an additional term $(\bk_2-\bk_1)$. Utilizing the coordinate
system of the unrotated reciprocal lattice we may write
write $\bG_1 = m_1\bb_1 + m_2\bb_2$ and
$\bG_2 = n_1\bb_1 + n_2\bb_2$ where $\bem = (m_1,m_2)^T$
and $\bnn = (n_1,n_2)^T$ must be integer valued.
Furthermore, the term $(\bk_2-\bk_1)$ is integer valued
in the coordinate system of the bilayer reciprocal lattice,
i.e., $(\bk_2-\bk_1) = l_1\bg_1 + l_2\bg_2$. The
transformation from the bilayer to unrotated reciprocal
lattice coordinate systems is

\begin{equation}
\bT_{BU} = \frac{\gamma}{i_3}
 \begin{pmatrix} -p+q & -2q \\ 2q & -p-q \end{pmatrix}
\end{equation}
and the rotation operator, transformed to the
unrotated reciprocal lattice coordinate system, is be found to
be

\begin{equation}
\bR_L = \frac{1}{i_3}
\begin{pmatrix} i_2+i_1 & -2i_1 \\
                2i_1 & i_2-i_1 \end{pmatrix}
\end{equation}
where, as before, we have

\begin{eqnarray}
i_1 & = & 2pq \\
i_2 & = & 3q^2-p^2 \\
i_3 & = & 3q^2+p^2.
\end{eqnarray}
Using these we may rewrite Eq.~(\ref{eq:Gcom}) as

\begin{equation} \label{eq:Gdio}
\begin{pmatrix} m_1 \\ m_2 \end{pmatrix} =
 \frac{1}{i_3} \begin{pmatrix} i_2+i_1 & -2 i_1 \\ 2i_1 & i_2-i_1 \end{pmatrix}
 \begin{pmatrix} n_1 \\ n_2 \end{pmatrix}
 + \frac{\gamma}{i_3} \begin{pmatrix} -p+q & -2q \\ 2q & -p-q \end{pmatrix}
 \begin{pmatrix} l_1 \\ l_2 \end{pmatrix}
\end{equation}
Our solution of this equation is based on diagonalising
$\bR_L$ and $\bT_{BU}$ which, it turns out, may be simultaneously
diagonalised. We find the eigenvalues of $\bR_L$ to be

\begin{equation}
a_\pm = -\frac{p\pm i\sqrt{3}q}{p\mp i\sqrt{3}q}
\end{equation}
and those of $\bT_{BU}$ to be

\begin{equation}
b_\pm = \frac{-\gamma}{p\mp i\sqrt{3}q}
\end{equation}
The eigenvectors in \emph{both cases} are given by

\begin{equation}
u_\pm = \frac{1}{\sqrt{2}} \begin{pmatrix} \frac{1}{2}(1\mp i\sqrt{3}) \\ 1 \end{pmatrix}
\end{equation}
Using these results we may rewrite Eq.~(\ref{eq:Gdio}) as

\begin{equation}
\bU^{-1}\bem = (\bU^{-1} \bR_L \bU)\bU^{-1}\bnn + (\bU^{-1} \bT_{CU} \bU) \bU^{-1} \bl
\end{equation}
Equating the real and imaginary parts of this equation we find

\begin{eqnarray}
(n_2+m_2)p & = & ((2n_1-n_2)-(2m_1-m_2))q - \gamma l_2 \\
((2n_1-n_2)+(2m_1-m_2))p & = & (m_2-n_2)3q - \gamma(2l_1-l_2)
\end{eqnarray}
and by introducing the new variables 
$n_3 = 2n_1-n_2$ and $m_3 = 2m_1-m_2$ we
can recast these equations as a
Diophantine problem:

\begin{eqnarray}
(m_2 + n_2)p & = & (n_3-m_3)q - \gamma l_2 \\
(n_3 + m_3)p & = & (m_2 - n_2)3q - \gamma (2l_1-l_2)
\end{eqnarray}
By absorbing the terms involving $\gamma$ either in
the coefficient of $p$ or $q$ these equations may
now be solved by inspection giving

\begin{eqnarray}
\left(m_2 + n_2 + \frac{\gamma l_2}{p}\right)p & = & (n_3-m_3)q \\
\left(n_3 + m_3 + \frac{\gamma (2l_1-l_2)}{p}\right)p & = & (m_2 - n_2)3q
\end{eqnarray}
which leads to the solutions

\begin{equation}
\bnn = \alpha \frac{1}{\gamma} \begin{pmatrix} p+q \\ 2p \end{pmatrix}
 + \beta \frac{1}{\gamma}\begin{pmatrix} 2q \\ -p+q \end{pmatrix} 
 - \frac{\gamma}{2p} \begin{pmatrix} l_1 \\ l_2 \end{pmatrix}
\end{equation}
and

\begin{equation}
\bem = \alpha \frac{1}{\gamma}\begin{pmatrix} -p+q \\ 2q \end{pmatrix}
 + \beta \frac{1}{\gamma}\begin{pmatrix} 2q \\ p+q \end{pmatrix}
 - \frac{\gamma}{2p} \begin{pmatrix} l_1 \\ l_2 \end{pmatrix}
\end{equation}
with $\alpha$ and $\beta$ integers. Note that we have immediately
written down the solution in terms of primitive vectors;
which may be done in a similar fashion to the real space
case (Appendix A).

Alternatively we may write

\begin{eqnarray}
(m_2 + n_2)p & = & \left(n_3-m_3-\frac{\gamma l_2}{q}\right)q \\
(n_3 + m_3)p & = & \left(m_2 - n_2 - \frac{\gamma(2l_1-l_2)}{q}\right)3q
\end{eqnarray}
which in turn leads to the solutions

\begin{equation}
\bnn = \alpha \frac{1}{\gamma}\begin{pmatrix} p+q \\ 2q \end{pmatrix}
 + \beta \frac{1}{\gamma}\begin{pmatrix} 2q \\ -p+q \end{pmatrix}
 - \frac{\gamma}{6q} \begin{pmatrix} l_1 - 2l_2 \\ 2l_1-l_2 \end{pmatrix}
\end{equation}
and

\begin{equation}
\bem = \alpha \frac{1}{\gamma}\begin{pmatrix} -p+q \\ 2q \end{pmatrix}
 + \beta \frac{1}{\gamma}\begin{pmatrix} 2q \\ p+q \end{pmatrix}
 + \frac{\gamma}{6p} \begin{pmatrix} l_1-2l_2 \\ 2l_1-l_2 \end{pmatrix}
\end{equation}

The case where $\delta=1$ proceeds in exactly the same manner,
the only difference being a different form for the
transformation matrix $\bT_{BU}$. The solutions are then found
to be

\begin{equation}
\bnn = \alpha \frac{1}{\gamma}\begin{pmatrix} -p+3q \\ -2p \end{pmatrix}
 + \beta \frac{1}{\gamma}\begin{pmatrix} 2p \\ p+3q \end{pmatrix}
 - \frac{\gamma}{6q} \begin{pmatrix} l_1 \\ l_2 \end{pmatrix}
\end{equation}

\begin{equation}
\bem = \alpha \frac{1}{\gamma}\begin{pmatrix} p + 3q \\ 2p \end{pmatrix}
 + \beta \frac{1}{\gamma}\begin{pmatrix} -2p \\ -p+3q \end{pmatrix}
 + \frac{\gamma}{6q} \begin{pmatrix} l_1 \\ l_2 \end{pmatrix}
\end{equation}
and

\begin{equation}
\bnn = \alpha \frac{1}{\gamma}\begin{pmatrix} -p+3q \\ -2p \end{pmatrix}
 + \beta \frac{1}{\gamma}\begin{pmatrix} 2p \\ p+3q \end{pmatrix}
 + \frac{\gamma}{6p} \begin{pmatrix} l_1 - 2l_2 \\ 2l_1-l_2 \end{pmatrix}
\end{equation}

\begin{equation}
\bem = \alpha \frac{1}{\gamma}\begin{pmatrix} p+3q \\ 2p \end{pmatrix}
 + \beta \frac{1}{\gamma}\begin{pmatrix} -2p \\ -p+3q \end{pmatrix}
 + \frac{\gamma}{6p} \begin{pmatrix} l_1-2l_2 \\ 2l_1-l_2 \end{pmatrix}
\end{equation}


\end{document}